\documentclass[aps,prb,twocolumn,showpacs,floatfix,superscriptaddress]{revtex4-1}

\usepackage{graphicx,color}
\usepackage{amsfonts}
\usepackage[figuresright]{rotating}
\usepackage{amssymb}
\usepackage{amsmath}
\usepackage{psfrag}
\usepackage{subfigure}
\usepackage{multirow}
\usepackage{tabularx}
\usepackage{textcomp}
\usepackage{units}
\usepackage{hyperref}
\usepackage{longtable}
\hypersetup{
 pdfnewwindow=true, colorlinks=true,
 linkcolor=blue, anchorcolor=blue,
 citecolor=blue, filecolor=blue,
 menucolor=blue, urlcolor=blue}

\def\nn{\nonumber}

\def\beq{\begin{eqnarray}}
\def\eeq{\end{eqnarray}}

\renewcommand{\v}[1]{\ensuremath{\mathbf{#1}}} 

\let\baraccent=\= 
\renewcommand{\=}[1]{\stackrel{#1}{=}} 
	
\makeatletter

\begin{document}
\title{Injection current in ferroelectric group-IV monochalcogenide monolayers}

\author{Suman\ \surname{Raj Panday}}
\affiliation{Department of Physics, Kent State University, Kent, Ohio 44242, USA}

\author{Salvador\ \surname{Barraza-Lopez}}
\affiliation{Department of Physics, University of Arkansas, Fayetteville, Arkansas 72701, USA}

\author{Tonatiuh\ \surname{Rangel}}
\affiliation{Department of Physics, University of California, Berkeley, California 94720, USA}

\author{Benjamin\ \surname{M. Fregoso}}
\affiliation{Department of Physics, Kent State University, Kent, Ohio 44242, USA}

\begin{abstract}
We study the injection current response tensor (also known as circular photogalvanic effect or ballistic current) in ferrolectric monolayer GeS, GeSe, SnS, and SnSe. We find that the injection current is perpendicular to the spontaneous in-plane polarization and could reach peak (bulk) values of the order of $10^{10}$A/V$^{2}$s in the visible spectrum. The magnitude of the injection current is the largest reported in the literature to date for a two dimensional material. To rationalize the large injection current, we correlate the injection current spectrum with the joint density of states, electric polarization, strain, etc. We find that various factors such as anisotropy, in-plane polarization and wave function delocalization are important in determining the injection current tensor in these materials. We also find that compression along the polar axis can increase the injection current (or change its sign), and hence strain can be an effective control knob for their nonlinear optical response. Conversely, the injection current can be a sensitive probe of the crystal structure.    
\end{abstract}

\maketitle
\section{Introduction}
\label{sec:intro}
Understanding the nonlinear optical response of materials is of great theoretical and experimental importance. The best known example of a nonlinear response is the second harmonic generation which is routinely used as a probe of symmetry (or lack of) or as a frequency multiplier in optoelectronic applications to mention two examples.~\cite{Boyd2008} Less well known is perhaps the so-called bulk photovoltaic effect (BPVE)\cite{Sturman1992} which refers to the generation of dc current in illuminated insulators lacking inversion symmetry. The BPVE has two components, namely, the injection current and the shift current, both quadratic in the optical electric field~\cite{Sturman1992,Paillard2018,Rioux2012,Baltz1981,Sipe2000,Morimoto2016,Spanier2016,Burger2019,Nakamura2017,Ogawa2017,Nagaosa2017,  Fregoso2017,Rangel2017,Wang2017b,Kushnir2017,Nakamura2018,Gong2018,Kushnir2019,Sotome2019,Sotome2019a,Kral2000,Cook2017,Zhang2019,Ibanez-Azpiroz2018,Brehm2018,Hosur2011,Juan2017,Rees,Chan2017,Flicker2018,Parker2019,Tan2019,Barik}. The injection current, in particular, has attracted attention for its role in the photovoltaic effect of ferroelectric materials~\cite{Sturman1992} including solar cell applications~\cite{Paillard2018}, in the coherent current control~\cite{Rioux2012}, and recently for its role as a probe of the topology of materials~\cite{Hosur2011,Juan2017,Rees,Chan2017,Flicker2018}.

The key features of the injection current (and of shift current) are as follows. First, it is generated in homogeneous materials. This should be contrasted with the working principle of standard semiconductor solar cells whose active region is a heterogeneous $pn$-junction. Second, the generated photovoltage can be many times larger than the energy band gaps, which means that carriers do not thermalize before they are collected (they are `hot'). This should be contrasted with the thermalization that takes place in $pn$-junctions much before carriers are collected which limits the photovoltage to a maximum given by the material's electronic band gap. Third, it depends on the polarization of light, specifically, the injection current vanishes for linear polarization of light and is maximum for circularly polarized light (hence the name circular photogalvanic effect).~\cite{Paillard2018}

The lack of inversion symmetry manifests in two distinct scenarios. In one scenario, the injection current is generated by photoexcited carriers which experience asymmetric momentum relaxation in the $\pm$\v{k} directions leading to a polar distribution and a net current~\cite{Sturman1992}. The origin of such asymmetric relaxation could be phonons, impurities, etc. In this scenario, the derivation of the injection current starts from a kinetic equation. In the second scenario, the injection current originates form photoexcited carriers pumped into $\pm$\v{k} of the Brillouin zone (BZ) at different \textit{rates} leading to a polar distribution and hence to a nonvanishing \textit{rate} of change of a charge current~\cite{Sturman1992,Sipe2000}. In this scenario the origin of the injection current is light-matter interactions not momentum relaxation. In this article we study the injection current arising from the light-matter interaction and add momentum relaxation phenomenologically.

Although progress in ferroelectric-based solar cells has been made~\cite{Spanier2016,Burger2019,Nakamura2017,Ogawa2017}, the BPVE photocurrents in bulk ferroelectrics are still small, resulting in smaller efficiencies compared with conventional Si-based cells. The experimental realization of the first two dimensional (2D) ferroelectric material~\cite{Chang2016} has led to a renewed interest in the BPVE in low dimensional systems.~\cite{Kral2000,Cook2017,Zhang2019} Of great interest are the group-IV monochalcogenide monolayer GeS, GeSe, SnS and SnSe predicted to be 2D ferroelectric materials below a critical temperature~\cite{Chang2016,mm2,mm3} with a large in-plane polarization. They offer a new (and simpler) platform to study the BPVE because they have simpler crystal structures than bulk ferroelectrics while having novel mechanical and optical properties including energy gaps in the visible range~\cite{Gomes2015,Naumis2017,Hu2019}.

In bulk ferroelectrics the dominant effect is believed to be the shift current~\cite{Young2012} but it is unknown if this holds true in 2D ferroelectrics. The shift current tensor in monochalcogenide monolayer GeS, GeSe, SnS and SnSe is beginning to be studied theoretically~\cite{Cook2017,Fregoso2017,Rangel2017,Wang2017b,Ibanez-Azpiroz2018} and experimentally, e.g., by the group of L. Titova~\cite{Kushnir2017,Kushnir2019}. It is predicted to be the largest reported so far in the literature for a two dimensional material and the largest reported for a three-dimensional (3D) material, if converted into an equivalent bulk value. The injection current tensor, however, has not been studied before in these materials. In this work we present a detailed study of the injection current in monochalcogenide monolayer GeS, GeSe, SnS and SnSe showing that both components of the BPVE are important.

\begin{figure}[]
\subfigure{\includegraphics[width=.45\textwidth]{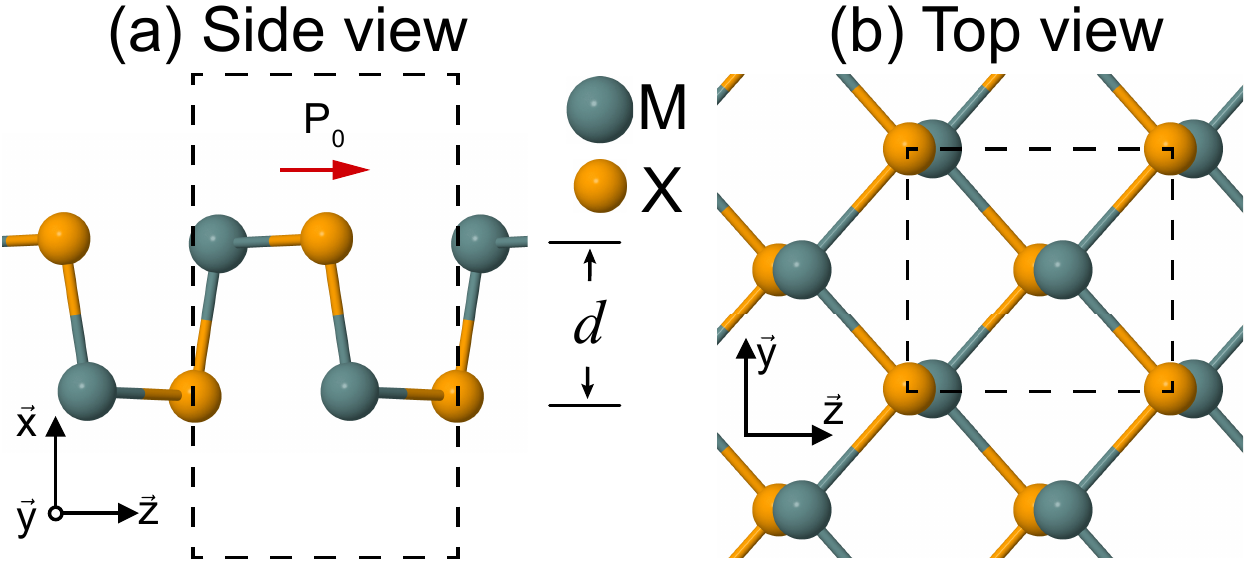}}
\caption{(a) Structure of monochalcogenide monolayers MX, where M=Ge,Sn X=S,Se. (b) The unit cell contains 4 atoms and is indicated in the boxes. The point group lacks an inversion center and develops a nonzero in-plane spontaneous polarization $P_0\hat{\v{z}}$.}
\label{fig:structure}
\end{figure}
Naively the injection current is expected to be proportional to the joint density of states (JDOS), with the idea that the more states available the larger the current. On the other hand, the injection current is not expected to correlate with the spontaneous polarization, because electric polarization is a ground state property whereas injection current involves excited states. This is supported by the heuristic argument that, contrary to shift current processes where there is an intrinsic length scale, injection current processes do not introduce a length scale which could naively be associated with some sort of `microscopic dipole'.
   
In this work, we test these ideas using analytic and numerical methods.  We find, to our surprise, that the injection current  tensor, hereafter called $\eta_2$, can reach peak (effective) values of the order $10^{11}$ A/V$^2$s, at photon energies corresponding to the visible light spectrum. This value is the largest reported so far in the literature, establishing the potential of these materials for optoelectronic applications. Contrary to expectations, the JDOS does not play a significant role. Rather, we find that a combination of factors including reduced dimensionality, in-plane polarization, and anisotropy can explain the large value of $\eta_2$ calculated in these materials.  

This paper is organized as follows. In Sec.~\ref{sec:methods} we describe the numerical procedure. In Sec.~\ref{sec:num_results} we calculate the injection current tensor $\eta_2$ as a function of photon energy and compare it with the JDOS. The correlation between polarization and injection current is presented in Sec.~\ref{sec:role_of_pol}. In Sec.~\ref{sec:2b_model} we introduce a simple two-band model for $\eta_2$ in GeS and conclude in Sec.~\ref{sec:conclusions}.

\section{Numerical methods}
\label{sec:methods}

We use density functional theory (DFT) as implemented in the ABINIT~\cite{Gonze2009} computer package, with the generalized gradient approximation to the exchange correlation energy functional as implemented by Perdew, Burke and Ernzerhof.\cite{Perdew1996} Hartwigsen-Goedecker-Hutter norm conserving pseudo potentials~\cite{Hartwigsen1998} were employed. To expand the plane waves basis set, energy cutoffs of 50 Hartree were employed for GeS and GeSe, and 60 Hartree for SnS and SnSe. The lattice parameter in the $x$-direction is set to 15 \AA~ (see Fig.~\ref{fig:structure}), which makes for more than 10 \AA~ of vacuum between slabs. To calculate the injection current tensor $\eta_2$, we included 20 valence and 30 conduction bands for GeS and SnS, and 30 valence and 20 conduction bands for GeSe and SnSe. They account for all allowed transitions up to 6 eV. 

To facilitate comparison of $\eta_2$ in monolayers with bulk materials we quote the monolayer $\eta_2$ as bulk (effective) value, i.e., assuming a (3D) stack of slabs (with the same orientation) comprising a bulk structure. To extract the effective response of a single layer, we scale the numerical result by the factor $\mathcal{L}/d$, where $\mathcal{L}$ is the supercell lattice parameter perpendicular to the slab, and $d$ is the effective thickness of the monolayer shown in Fig.~\ref{fig:structure}. For concreteness, we estimate the slab thicknesses as 2.56, 2.59, 2.85 and 2.76 \AA~ for GeS, GeSe, SnS, and SnSe, respectively.  Once the ground-state wave function and energies were computed, the TINIBA package~\cite{tiniba} was used to compute $\eta_2$ as implemented  in Ref.~\onlinecite{Sipe2000}. The sum over $\v{k}$-points is made using the interpolation tetrahedron method~\cite{bloch-tetra}. See the Appendices for more details.

\section{Numerical results}
\label{sec:num_results}
\begin{figure}[]
\subfigure{\includegraphics[width=.45\textwidth]{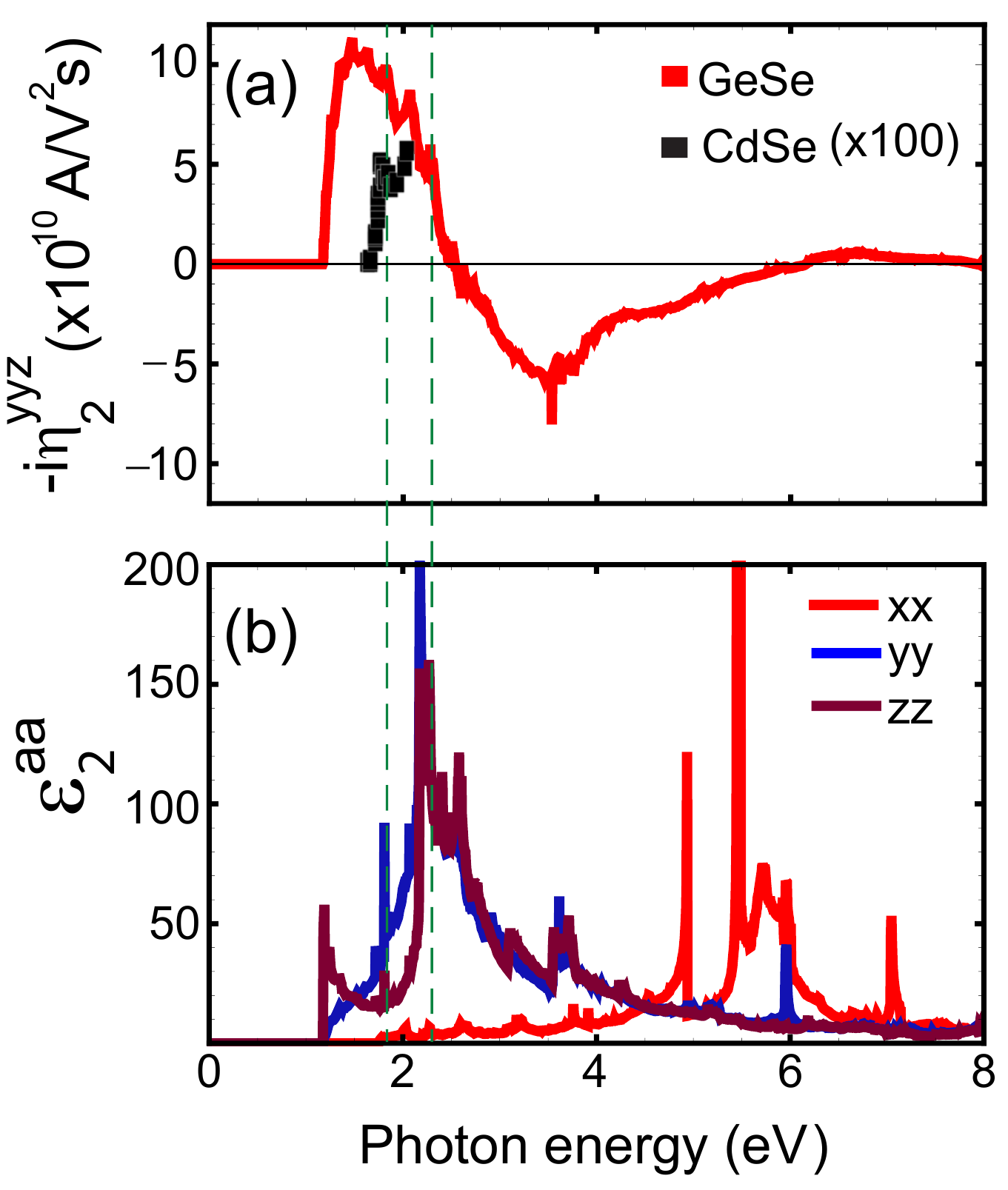}}
\caption{(a) Effective injection current susceptibility tensor of single-layer GeSe (red line) and CdSe (black squares)~\cite{Laman1999}. The injection current of GeSe is two orders of magnitude larger than that of CdSe and peaks for photon energies in the visible spectrum. The large magnitude of the injection current in the visible light spectrum highlights the potential of these materials for  optoelectronic applications. (b) Peaks in the imaginary part of the dielectric function, Eq.~\ref{eq:epsilon2}, shows no obvious correlation with injection current.}
\label{fig:inj_curr_GeSe}
\end{figure}

\subsection{Injection current}
\label{sec:inj_current tensor}
An incident monochromatic optical field $E^a = E^{a}(\omega)e^{-i\omega t}$ + c.c. induces an injection current governed by the equation
\begin{align}
\frac{d}{dt}J^{a}_{inj} = 2{\eta_{2}}^{abc}E^{b}({\omega})E^{c}({-\omega})-\frac{J^{a}_{inj}}{\tau},
\label{eq:injection_tau}
\end{align}
where summation over repeated indices is implied, $\tau$ is a phenomenological momentum relaxation time, and $\eta_{2}^{abc}(0,\omega,-\omega)$ is the injection current tensor~\cite{Sturman1992, Sipe2000}
\begin{align}
 \eta_{2}^{abc} = \frac{e^{3}\pi}{2{\hbar}^{2}V } \sum_{nm\v{k}} \omega_{mn,a} f_{nm}[r^{c}_{mn},r^{b}_{nm}]{\delta}({\omega}_{mn}-{\omega}).
\label{eq:eta2}
\end{align}
Here $e=-|e|$ is the charge of the electron, $a,b,c=x,y,z$ are Cartesian components, $n$ is a band index, $f_{nm}=f_n-f_m$  is the difference in occupation numbers at zero temperature of bands $n$ and $m$, $\hbar\omega_n$ is the energy of the band $n$, $r^{a}_{nm}=i\langle u_{n}|u_{m,a}\rangle$ is the the Berry connection, $u_{n}$ is the periodic part of the Bloch wavefunction, and $[r^{c}_{mn},r^{b}_{nm}]\equiv r^{c}_{mn}r^{b}_{nm} - r^{b}_{mn}r^{c}_{nm}$. We define the subscript $X_{,a}$ to mean derivative with respect to the crystal momentum with Cartesian coordinate $a$. For example, $\omega_{nm,a}\equiv (\omega_{n}-\omega_{m})_{,a} = \omega_{n,a} - \omega_{m,a} = v_{n}^a - v_{m}^a$ are band-velocity differences. In the thermodynamic limit in $d$-dimensions we have $(1/V)\sum_{\v{k}} \to \int d\v{k}/(2\pi)^d$, where the integral is over the (BZ) and $V$ is the sample volume.

\begin{table}
\caption{Peak values of $\eta_2$ reported in representative 2D and 3D materials. The photon energy, spontaneous polarization $|\v{P}_0|=P_0$, and theoretical (th.) vs experimental (exp.) values are indicated. MX stands for monolayer of GeS, GeSe, SnS or SnSe. For 2D materials the effective value of $\eta_2$ is reported. GaAs has zero $\eta_2$ by symmetry.}
\begin{center}
 \begin{tabular}{|c|c|c|c|c|}
 \hline
  Material  & $|\eta_2|$              & $\hbar\omega$         & $P_0$                  & ~~Ref.~~ \\
             & ($\times 10^8$ A/V$^2$s) & ~~(eV)~~              & ($\mu$C/cm$^2$)        &          \\
 \hline
  MX (2D)                        & 100-1000              & 1.5-2.5               & 72-195~[\onlinecite{Rangel2017}]    &  present \\
 \hline
  CdSe (3D)      & 7                        & 2.2                   & 0.6~[\onlinecite{Schmidt1997}]   &  \onlinecite{Nastos2010}(th.)\\
 \hline
  CdSe (3D)      & 5                        & 2                     &   0.6~[\onlinecite{Schmidt1997}]  & \onlinecite{Laman1999}(exp.) \\
 \hline 
  CdS (3D)       & 4                        & 2.8                   &                                   & \onlinecite{Nastos2010}(th.) \\
 \hline
  CdS (3D)       & 4                        & 3                     &                        &   ~\onlinecite{Laman2005}(exp.)\\
 \hline
 CdSe (3D)       & 1.5                      & 1.8                   &   0.6~[\onlinecite{Schmidt1997}]  & \onlinecite{Laman2005}(exp.) \\
 \hline
  MoS$_2$(2D)    & 10$^{-7}$                & 2.8                   &  0   &  \onlinecite{Arzate2016}(th.)\\
 \hline
	GaAs (3D)      & 0                        &                       & 0                      & \\
	\hline
\end{tabular}
\end{center}
\label{table:eta2_literature}
\end{table}
It can be shown that $\eta_2^{abc}$ is a pure complex third rank tensor which is antisymmetric in the last two indices~\cite{Sturman1992}. For this reason, it vanishes for linearly polarized light and is maximum for circularly polarized light.

Monolayers of GeS, GeSe, SnS and SnSe have $mm2$ point group which contains a two-fold (polar) axis, two mirror planes and lacks center of inversion. Accordingly~\cite{Boyd2008}, the nonzero components of $\eta_2^{abc}$ are $zxx$, $zyy$, $zzz$, $yyz$, $xzx$, $xxz$ and $yzy$. In addition, the antisymmetry of $\eta_2^{abc}$ with respect to exchange of the last two indices forces the $zxx$, $zyy$, $zzz$ components to vanish, leaving only two independent components, namely, $yyz$, $xxz$. Since the plane of the slab is perpendicular to the $x$-axis, see Fig.~\ref{fig:structure}, the component $xxz$ is much smaller than the $yyz$ component. Here we focus on the relevant $yyz$ component 
 
\begin{align}
 \eta_{2}^{yyz} = \frac{e^{3}\pi}{2{\hbar}^{2}V } \sum_{nm\v{k}} \omega_{mn,y} f_{nm} [r^{z}_{mn},r^{y}_{nm}]{\delta}({\omega}_{mn}-{\omega}).
\label{eq:eta2_yyz}
\end{align}
Note that this component fixes the $y$-direction of the current flow to be perpendicular to the spontaneous polarization $z$, $\v{P}_0=\hat{\v{z}}P_0$. In Fig.~\ref{fig:inj_curr_GeSe}(a) we show the representative spectrum of (effective) $\eta_2^{yyz}$ for monolayer GeSe as a function of incident photon energy $\hbar \omega$. The responses of monolayer GeS, SnS and SnSe are similar and presented in Appendix~\ref{sect:injection current}. Note that $\eta_2^{yyz}$ is zero for photon energy less than the energy gap and peaks in the visible light spectrum ($1.5-3$ eV) at about $10^{11}$ A/V$^{2}$s. As the photon energy increases, the injection current reverses its direction several times and progressively decreases in magnitude. 

To give perspective on this value of $\eta_2$, other values for different materials are shown in Table~\ref{table:eta2_literature}. Surprisingly, the peak value of $\eta_2$ in monolayer GeSe is two orders of magnitude larger than typical values and many orders of magnitude larger than monolayer MoS$_2$~\cite{Arzate2016}. This demonstrates the potential of monolayer GeS, GeSe, SnS and SnSe in  optoelectronic applications.

\subsection{Linear dielectric function}
\begin{figure}[]
\subfigure{\includegraphics[width=.48\textwidth]{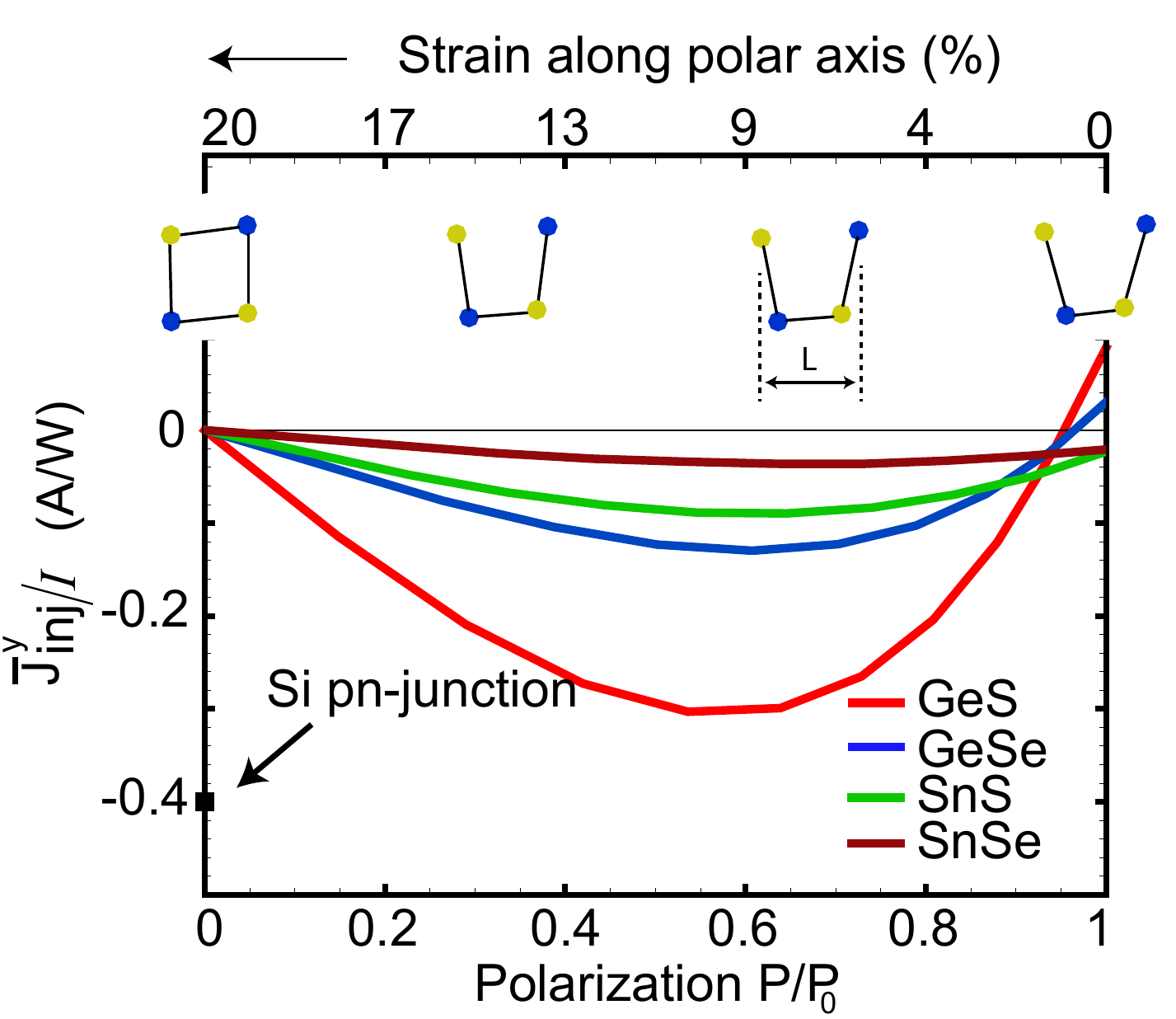}}
\caption{Average current per unit incident intensity (I) and the polarization (strain) along a path in configuration space for various materials. For a given material, the average current is nonmonotonic, with an optimal value of polarization (strain) yielding maximum photocurrent. On the other hand, the larger the \textit{spontaneous} polarization of a material ($P_0$) the larger the magnitude of the injection current. To calculate the current we use reasonable parameters for clean semiconductors $\tau=100$ fs~[\onlinecite{Li2018}], $\varepsilon/\varepsilon_0 \sim 4$~[\onlinecite{Gomes2015}], $W=10$ eV. For comparison, the state of the art Si-based solar cells produce a current of about 400 mA/W.[\onlinecite{Pagliaro2008}] $L$ is defined as $z$-distance between outmost atoms in the unit cell as shown in the inset.}
\label{fig:integral_eta2_polarization}
\end{figure}

An explanation for the large $\eta_2$ in these materials could be the large number of available states. The more states available, the larger the absorption of photons, and the larger the nonlinear optical response. To test this hypothesis we compare the spectrum of $\eta_2$ with that of the imaginary part of the linear dielectric function
\begin{equation}
\varepsilon_{2}^{ab}(\omega)=\frac{4e^{2}\pi^2}{\epsilon_{0}\hbar V}  \sum_{nm\v{k}} f_{nm} r^{a}_{nm}r^{b}_{mn}\delta{(\omega_{mn}-\omega)},
\label{eq:epsilon2}
\end{equation}
which roughly follows the JDOS. Our dielectric tensor $\varepsilon_{2}$ for GeSe is shown in Fig.~\ref{fig:inj_curr_GeSe}(b) and agrees with previous reports~\cite{Gomes2015,Rangel2017}. $\varepsilon_{2}$ is largest at energies near the van Hove peaks, so if light absorption is the origin of the large $\eta_2$ both spectra should peak at those same energies. Inspection of Fig.~\ref{fig:inj_curr_GeSe}(b) shows this is not the case in general. Only few peaks in $\eta_2$ match the energy locations of van Hove singularities indicated by dashed vertical lines. In addition, the average $\epsilon_2$ in these materials is not particularly large. This suggest the JDOS cannot, by itself, be the cause of the large magnitude of $\eta_2$.

\section{The role of polarization in $\eta_2$ in monolayer GeS, GeSe, SnS and SnSe}
\label{sec:role_of_pol}
We now investigate the role of electric polarization in $\eta_2$ in monolayer GeS, GeSe, SnS and SnSe. It is worth recalling that shift current processes introduce an intrinsic length scale (so-called shift vector) which naively could be associated with some sort of microscopic dipole. For this reason, one could expect that the larger the polarization of the material the larger the shift current. It has been shown that there is no general relation between polarization and shift current in 3D ferroelectrics~\cite{Tan2016}. In 2D ferroelectric materials, there may be a stronger correlation due to reduced dimensionality~\cite{Fregoso2017}.

The injection current, on the other hand, does not introduce an intrinsic length scale and hence we do not expect any correlation between polarization and $\eta_2$. Yet, the difference in $\eta_2$ between nonferroeletric monolayer MoS$_2$ and ferroelectric monolayer GeSe of several orders of magnitude (Table~\ref{table:eta2_literature}) suggests at least an indirect relation. 

In this section we attempt to test this idea by performing numerical experiments. We compare the average injection current {\em versus} the polarization of monolayer GeS, GeSe, SnS, and SnSe as we vary their atomic positions along a path in configuration space from the centrosymmetric to the noncentrosymmetric configuration. The ground state of monolayer GeS, GeSe, SnS, and SnSe is the noncentrosymmetric configuration with an spontaneous polarization $P_0$; the other metastable states could, in principle, be reached by applying a strain. 

\subsection{Average injection current}
From Eq.~\ref{eq:injection_tau}, a circularly polarized optical field induces a current given by
\begin{align}
\frac{d}{dt}J_{inj}^y = \pm i\eta_{2}^{yyz} |E_0|^2 - J_{inj}^y/\tau,
\end{align}
where $\pm$ determines the chirality of light and $E_0$ is the amplitude of the optical field. Importantly, note that the current is perpendicular to the polarization axis. We used the convention in which $E^{a}(\omega)=E^{a}_0/2$  [\onlinecite{Boyd2008}]. In steady state, the injection current is $J_{inj}^y = \pm \tau i\eta_{2}^{yyz} |E_0|^2$. If the light has a broad spectrum the frequency average of the current is a good measure of the overall injection current generated 
\begin{align}
\bar{J}_{inj}^y = \frac{1}{\Delta\Omega}\int_{E_g}^{W} d\omega J_{inj}^y,
\label{eq:Javg}
\end{align}
where $\hbar\Delta\Omega=(W-E_g)$, $W$ is the bandwidth of the incident light, and $E_g$ is the energy band gap of the material. 

\subsection{Electric polarization}
We compute the electric polarization for each material at various atomic positions using the modern theory of polarization~\cite{King-Smith1993,Resta1994} as implemented in ABINIT. We first identify a smooth path in configuration space between the ground state and the centrosymmetric geometry with zero polarization. The atomic displacements along a (one-dimensional) path are parametrized by $\lambda$ as $\v{R}^i(\lambda)=\v{R}^i_0 + \lambda (\v{R}^i_f-\v{R}^i_0)$, where $\v{R}^i_0$ ($\v{R}^i_f$) is the initial (final) position of atom $i$th atom in the centrosymmetric (noncentrosymmetric) structure. The  minimum-energy path between the $\pm \v{P}_0$ configurations~\cite{Rangel2017,Wang2017} is indistinguishable from the linear path used here. The electric polarization has ionic and electronic components

\begin{equation}
\v{P}(\lambda) = \frac{e}{V}\sum_{i}Z^{i}\v{r}_{i}(\lambda)-\frac{e}{V} \sum_{v\v{k}}  \pmb{\xi}_{vv}(\lambda),
\label{eq:pol_formula}
\end{equation}
represented by the first and second terms above. $\v{r}_i$ and $Z^{i}$ are the position and atomic number of the $ith$ ion, $V$  is the simulation volume, and $\pmb{\xi}_{vv}(\lambda)=i\langle u_{v}^{\lambda} | u_{v,a}^{\lambda} \rangle$ is the Berry connection of the valence band $v$. Summation is over occupied states. The details of the calculation are presented in Appendix~\ref{sect:polarization_and_inj}. Note that the motion of the atoms occurs only along the $z$-direction (Table~\ref{table:coords} in Appendix~\ref{sect:polarization_and_inj}), which is equivalent to straining the lattice along the polar axis. To quantify the strain we define 

\begin{align}
strain= \frac{L- L_0}{L_0},
\end{align} 
where $L$ is defined in Fig.~\ref{fig:integral_eta2_polarization}, and $L_0$ the value in the ground state.

\subsection{Comparison of injection current and polarization}
\begin{figure}
\subfigure{\includegraphics[width=.35\textwidth]{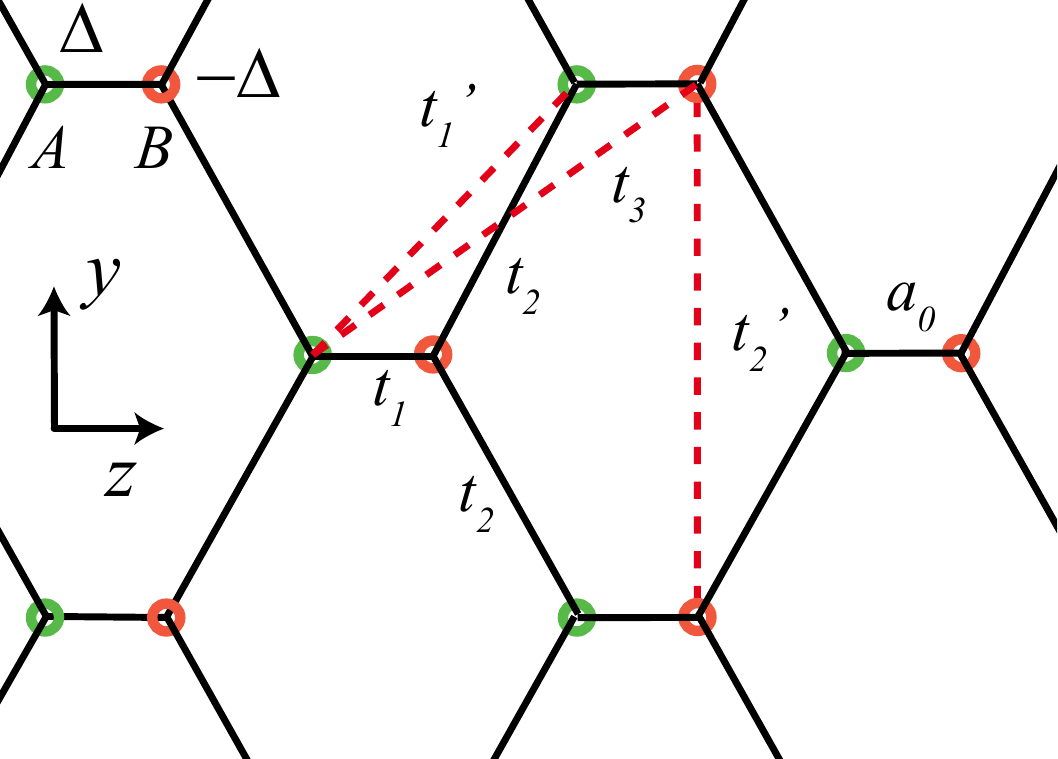}}
\caption{Tight-binding (TB) model of monolayer GeS near the band edge. The lattice has two nonequivalent sites $(A,B)$ per unit cell. The hopping parameters and onsite potential are indicated. The crystal lacks inversion symmetry and develops spontaneous polarization along the $z$-axis. $a_0$ is the distance between the A, and B sites. The lattice vectors (not shown) are $\v{a}_1=(a_z,-a_y)$ and $\v{a}_2=(a_z,a_y)$. The details of the model are presented in Appendix~\ref{sec:tb_model}.}
\label{fig:tb_model}
\end{figure}

In the calculation of the injection current we assumed reasonable semiconductor parameters, $\tau=100$ fs[\onlinecite{Li2018}], $\varepsilon/\varepsilon_0 \sim 4$[\onlinecite{Gomes2015}], $W=10$ eV and used $I=\sqrt{\varepsilon/\mu}E_0^2/2$. Fig.~\ref{fig:integral_eta2_polarization} shows the average current as a function of polarization $\v{P}=\hat{\v{z}}P$ for various atomic configurations. In all materials, the average injection current is a nonmonotonic function of polarization (or strain). Its absolute value reaches a local maximum at an optimal polarization $0.6 P_0$ (or $\sim 9\%$ strain). Also, the current of single-layer GeSe and GeS can change sign even with a small strain, $\sim 1\%$, suggesting a new way to engineer the injection current. 

This interesting behavior of the average injection current indicates that at least two competing effects are at play. For example,  polarization increases injection current at small polarization but then at large polarizations the wave function overlaps (or lack thereof) decrease $\eta_2$ and even change its sign. In fact, wave function delocalization has been found to play an important role in shift current processes.~\cite{Tan2019} It is interesting to note that the polarization at which the magnitude of the average current is maximum is roughly the same for all materials. Similarly, monolayer GeS and GeSe can have zero average injection current whereas SnS and SnSe don't. This likely reflects the wavefunction differences between Ge and Sn atoms in this crystal structure.

An important point to note is that the magnitude of the average injection current of GeS (red) is the largest whereas that of SnSe (brown) is the smallest. This correlates with the magnitude of their spontaneous polarizations in the ground state. This ordering persist for metastable states where $P<P_0$ indicating a correlation of the injection current with the electric polarization.

\section{Two-band model of $\eta_2$ in monolayer GeS}
\label{sec:2b_model}

\begin{figure}
\subfigure{\includegraphics[width=.47\textwidth]{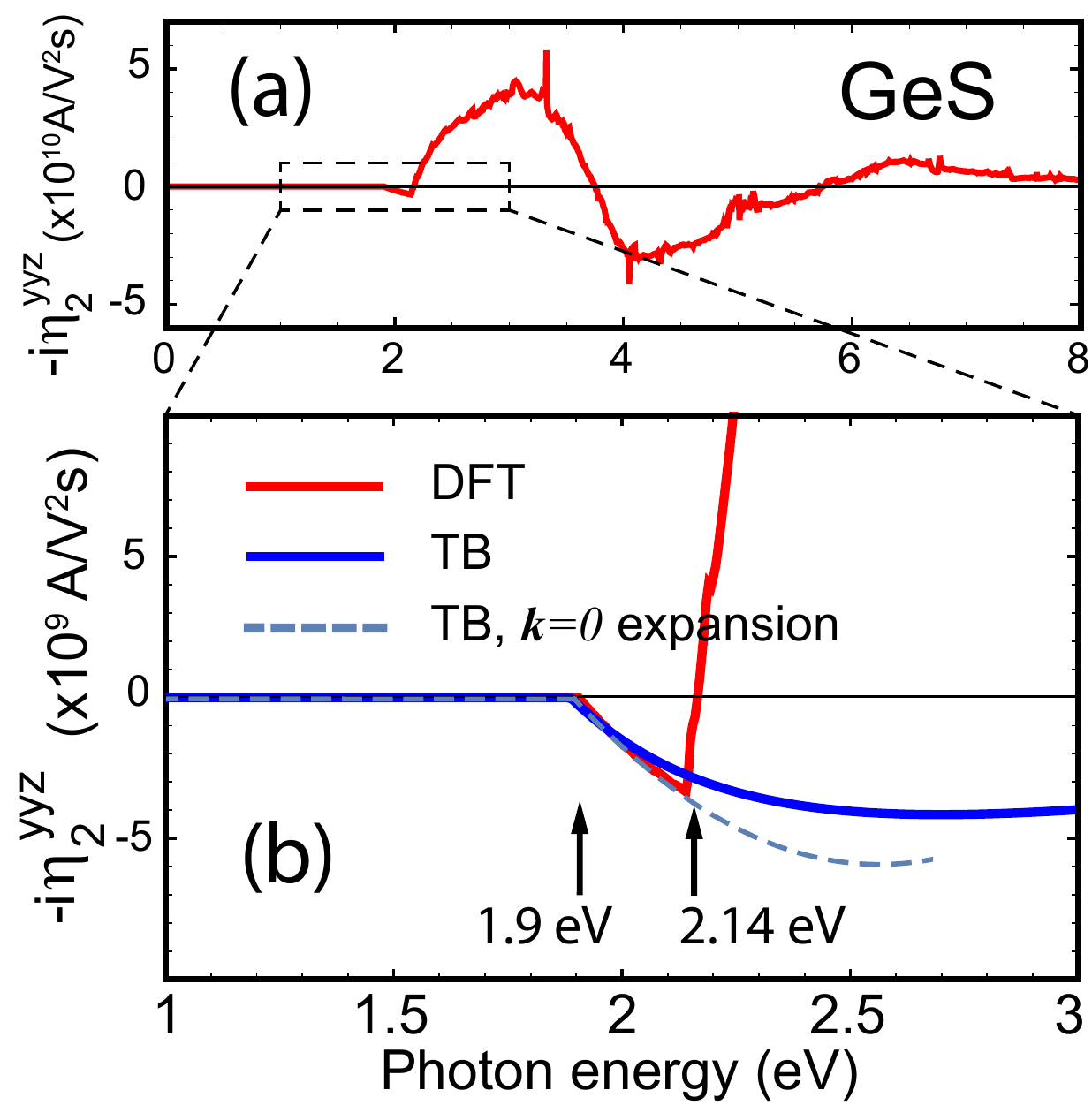}}
\caption{Effective injection current tensor of single-layer GeS from DFT (top panel) and tight-binding (TB) model (lower panel). The TB model reproduces the injection current from DFT near the band edge (near the $\Gamma$ point) in the energy range of (1.9, 2.14) eV. For photon energies larger than 2.14 eV other regions in the BZ contribute, see Fig~\ref{fig:ges_band_projection}. Dashed line shows the analytical result for a small momentum expansion about $\v{k}=0$, Eq.~\ref{qe:eta2_analitical_tb}.}
\label{fig:tb_injection}
\end{figure}

To disentangle the factors contribution to $\eta_2$, we construct a simple 2D two-band tight-binding model of monolayer GeS. Its hopping parameters and onsite potentials are shown in Fig.~\ref{fig:tb_model}. The model reproduces the calculated DFT shift current in this material near the band edge~\cite{Cook2017} and, importantly for our purposes, it can be solved analytically (near de band edge) helping disentangle the contributions to $\eta_2$ from JDOS, anisotropy, polarization, etc. As shown in Appendix~\ref{sec:tb_model} the response tensor of a two-band model can be written as~\cite{Hosur2011,Juan2017}

\begin{align}
\eta_{2}^{yyz} = \frac{i e^3 \pi}{2 \hbar^2 V}\sum_{\v{k}} \omega_{cv,y}\Omega^{x}_{c}\delta(\omega_{cv}-\omega),
\label{eq:eta2_2B}
\end{align} 
where $\Omega^{x}_c ~(=-\Omega_{v}^{x})$ is the Berry curvature of the conduction band. Fig.~\ref{fig:tb_injection}(b) shows $\eta_2$ obtained from the tight-binding model and its comparison with the DFT injection current for monolayer GeS. The agreement is good in the energy range between 1.9 and  2.14 eV. For photon energies above 2.14 eV other phototransitions in the BZ contribute, see Fig.~\ref{fig:ges_band_projection}(b). Near the band edge the energy bands an Berry curvature can be expanded in powers of the momenta giving 

\begin{align}
\hbar\omega_{cv} &= E_g + \alpha a_z^2 k_z^2 + \beta a_y^2 k_y^2 + \gamma a_z^2 a_y^2 k_z^2  k_y^2 + ... \nn \\
\Omega^{x}_c &= (A a_0 + B a_z) a_y^2 k_y + C a_y^2 a_z^2 k_y  k_z^2 +...,
\end{align}
where $E_g= 1.89$ eV is the energy band gap, ($\alpha,\beta,\gamma)=(2.30,1.33,-2.11)$ eV are constants that parametrize the curvature of the bands $(\alpha,\beta)$, $A=0.30, B=0.34$ are dimensionless constants that parametrize the Berry curvature near the band edge and depend on the hopping parameters, and $C=-1.237 a_0 - 1.272 a_z$. $(a_z,a_y,a_0,d)=(2.26,1.82,0.62,2.56)$~\AA~ are the lattice parameters $(a_z,a_y)$, the distance between the A-B sites $(a_0)$ and the thickness of the slab $d$, respectively. As shown in Appendix~\ref{sec:tb_model}, leading term in $\eta_2$ is 

\begin{align}
\eta_{2}^{yyz} &= \frac{i e^3 \pi}{2\hbar^2} \bigg(\frac{\bar{\Omega}}{2\pi}\frac{E_g}{\sqrt{\alpha\beta}}\bigg) \left(\frac{\hbar \omega- E_g}{E_g}\right) +\cdots. \nn \\
&~~~~~~~~~~~~~~~~~~~~~~~~~~~~~~~~~~~~~~~~~~~~\hbar\omega\geq E_g  
\label{eq:eta_2_expansion}
\end{align} 
We have written this equation in a form that emphasizes the different factors that contribute to $\eta_2$. The $\bar{\Omega}$ factor is dimensionless and comes from the Berry curvature and it is given by

\begin{align}
\bar{\Omega} = \frac{a_y (A a_0 + B a_z)}{a_z d} \approx 0.3
\end{align}
The Berry curvature has two contributions of order $a_y a_z$ and $a_y a_0$, respectively. The second contribution, which depends on $a_0$ could roughly be interpreted as coming from the polarization of the material. It is about 20\% of the total $\eta_2$ showing, again, that polarization does play a role in $\eta_2$. The other important dimensionless factor, $E_g/\sqrt{\alpha\beta}$, is roughly the geometric mean of the radius of curvature of the band along two directions at $\v{k}=0$. For monolayer GeS is given by 

\begin{align}
\frac{E_g}{\sqrt{\alpha\beta}} \approx 1.1
\end{align}
Finally, the constant prefactor is $i e^3 \pi/2\hbar^2 = -i6\times 10^{11}$ A/V$^2$s and hence $\eta_2$ near the band edge is

\begin{align}
-i\eta_{2}^{yyz} \approx - (2 \times 10^{10} A/V^2s) \left(\frac{\hbar \omega- E_g}{E_g}\right)+\cdots.
\end{align} 

\subsection{Three dimensions}
A similar calculation but assuming three dimensions gives the leading term near the band edge

\begin{align}
\eta_{2}^{yyz} &= \frac{i e^3 \pi}{2\hbar^2} \bigg(\frac{\bar{\Omega}}{4\pi^2}\frac{E_g^{3/2}}{\sqrt{\alpha\beta\beta'}}\bigg) \left(\frac{\hbar \omega- E_g}{E_g}\right)^{3/2} +\cdots, \nn \\
&~~~~~~~~~~~~~~~~~~~~~~~~~~~~~~~~~~~~~~~~~~~~\hbar\omega\geq E_g  
\label{eq:eta_2_expansion_3d}
\end{align} 
where $\beta'$ parametrizes the curvature along $x$, and Berry contribution is $\bar{\Omega} = a_y (A a_0 + B a_z + C' a_x)/a_z a_x$ ($C'$ is a constant). This shows that in 3D the rise of $\eta_2$ with frequency is slower than in 2D and suggests that reduced dimensionality is important for large $\eta_2$ in monolayer GeS. The simple model studied is only valid near the Gamma point. However anisotropy and reduced dimensionality are likely important at transitions at other points in the BZ.

\section{Conclusions}
\label{sec:conclusions}
Understanding the nonlinear optical response of novel materials is of fundamental importance for advancing the field of optoelectronics. Monolayer GeSe, GeS, SnS and SnSe are predicted to be ferroelectric materials with in-plane polarization and to exhibit promising electronic, mechanical and optical properties~\cite{Gomes2015,Naumis2017,Hu2019}. Here we studied the injection current tensor $\eta_2$ which is one of the contributions to the bulk photovoltaic effect in these materials. 

We find that the spectrum of $\eta_2$ peaks at (bulk) values near $10^{11}$ A/V$^2$s in the visible range and is the biggest reported so far. Contrary to bulk ferroelectric materials where the shift current is expected to dominate~\cite{Young2012}, in these materials both the shift current and injection current tensors are expected to be very large. Among our main results, we predict that the injection current can flow only perpendicular to the polar axis and that the injection current can increase or change sign upon compression along the polar axis of these materials.  

We showed that the JDOS, alone, cannot explain $\eta_2$ in these materials but rather a combination of factors including, in-plane polarization, reduced dimensionality, anisotropy, and covalent bonding (wave function delocalization, see Fig.~\ref{fig:ges_band_projection}) all of which are closely intertwined. If and how these factors enhance $\eta_2$ in other materials should be studied on a case by case basis. In summary, our results characterize the injection current tensor in monolayer GeS, GeSe, SnS and SnSe. Their relatively simple crystal structure and novel in-plane ferroelectricity make them an ideal playground for novel nonlinear optical phenomena. 

\textit{Note added}. Recently, similar results of injection current in monolayer GeS have been presented in Ref.~\onlinecite{Wang2019}.

\section{Acknowledgments}

BMF and SRP thank contract No. DE-AC02-05CH11231. SBL acknowledges the U.S. Department of Energy, Office of Basic Energy Sciences, Early Career Award DE-SC0016139.

\appendix

\section{Electronic band structures}
\label{sect:bandstructures}

\begin{table}
\caption{Band gaps and effective spontaneous polarization $\v{P}_0$ of monolayer GeS, GeSe, SnS and SnSe within DFT-PBE.}
\begin{center}
 \begin{tabular}{|c|c|c|c|}
 \hline
 \textbf{Monolayer} & \textbf{Direct gap} & \textbf{Indirect gap} & \textbf{Polarization}\\
  	 $  $ & $eV$ & $eV$ & $C/m^2$\\
 \hline
  GeS& 1.89 & 1.73 & 1.95\\
 \hline
  GeSe& 1.16 & 1.16 & 1.38\\
 \hline
  SnS& 1.57 & 1.46  &  0.95\\
 \hline
  SnSe & 0.95 & 0.95  & 0.72\\
 \hline
\end{tabular}
\end{center}
\label{table:E_g_MMs}
\end{table}

\begin{figure}[t]
\includegraphics[width=.43\textwidth]{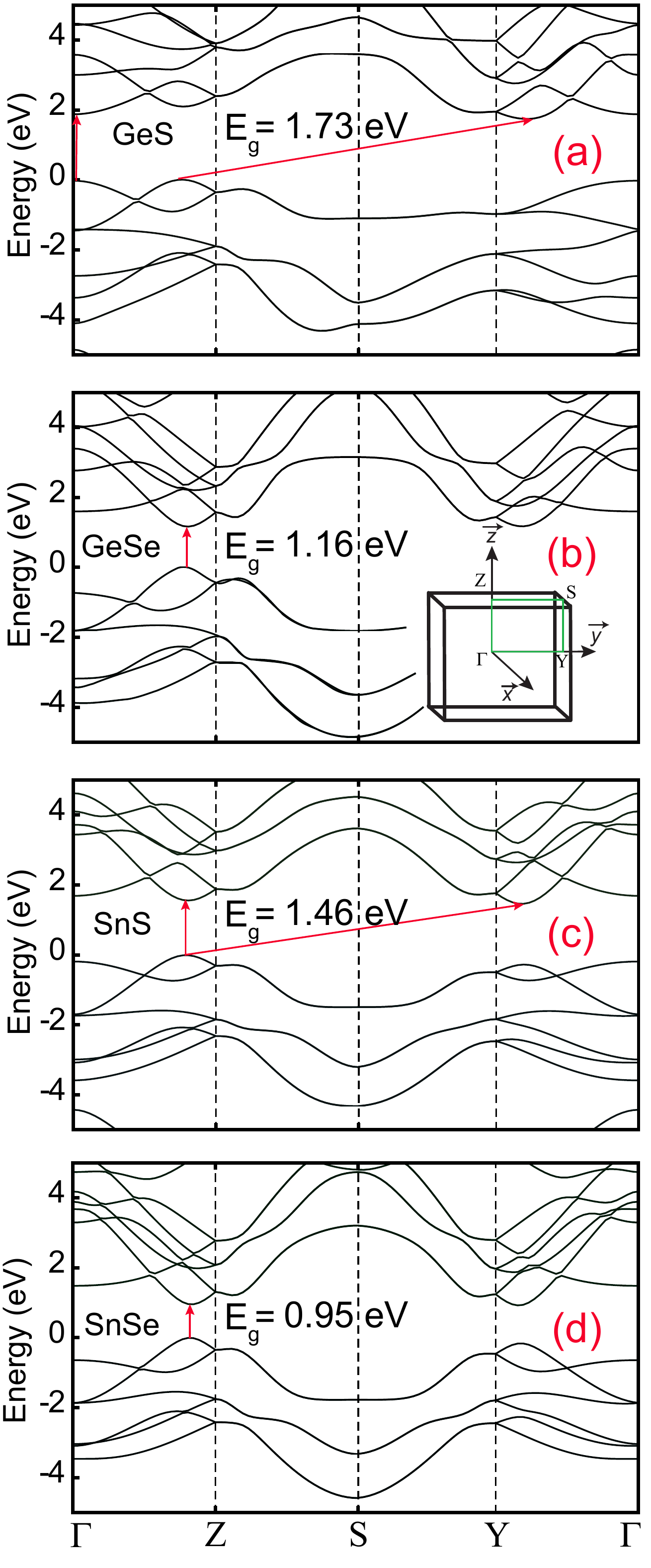}
\caption{Electronic bandstructure of monolayer GeS, GeSe, SnS and SnSe calculated within DFT-PBE. The inset indicates the path in the Brillouin zone (BZ). Red arrows indicate direct/indirect gaps. }
\label{fig:bandstructure}
\end{figure}

The electronic band structures of monolayer GeS, GeSe, SnS and SnSe calculated within DFT are shown in Fig.~\ref{fig:bandstructure}. They agree with previous works in the literature~\cite{Singh2014}. The bandgaps are also indicated in the figure and in Table~\ref{table:E_g_MMs}.

\section{Injection current and linear spectra}
\label{sect:injection current}
\begin{figure}[]
\includegraphics[width=.45\textwidth]{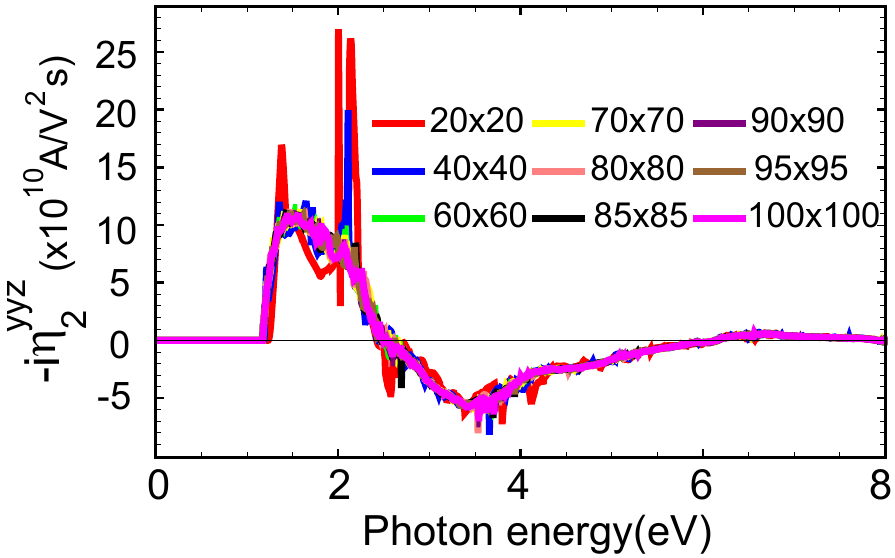}
\caption{Convergence of the injection current tensor for GeSe monolayer with respect to the $\v{k}$-point mesh size.}
\label{fig:convergence}
\end{figure}

\begin{figure*}[]
\includegraphics[width=\textwidth]{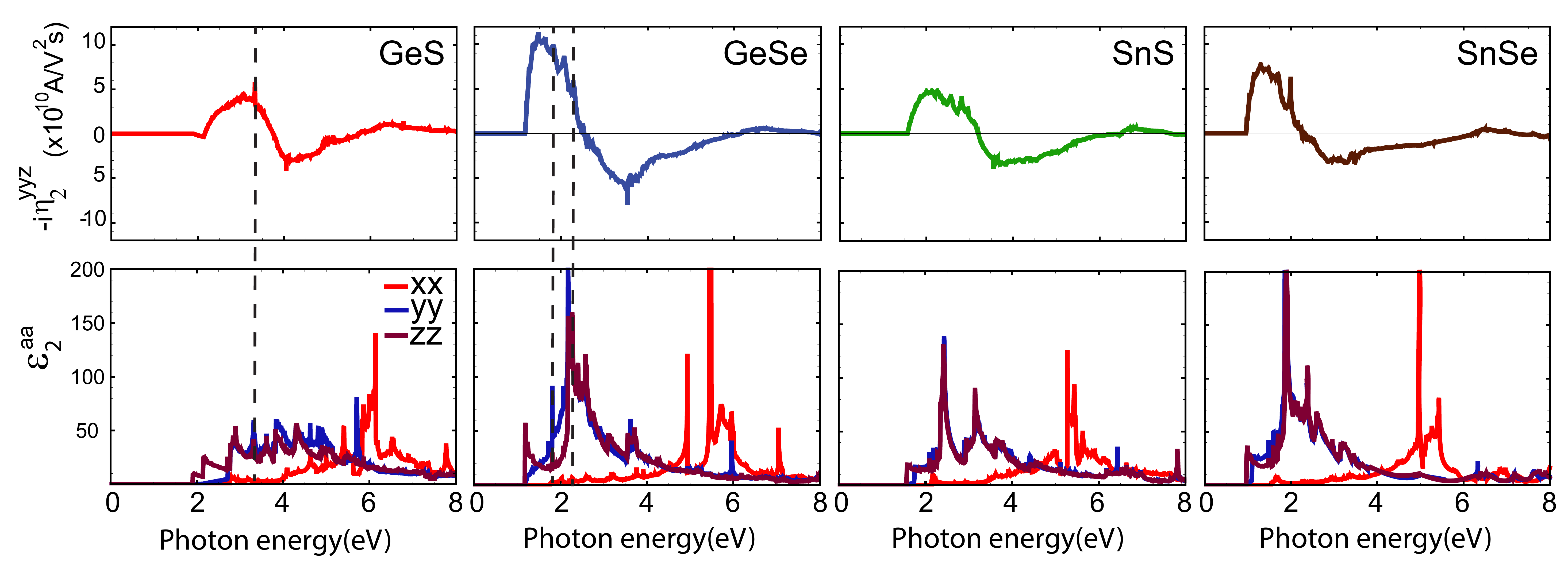}
\caption{ (top panels)  The injection current tensor of group-IV monochalcogenides monolayers. For comparison we also show the imaginary part of the linear dielectric function $\varepsilon_2$ (bottom panels) which determines the optical absorption of the material. As can be seen, there is no obvious correlation between them.}
\label{fig:inj_curr_all}
\end{figure*}

We used the TINIBA package~\cite{tiniba} to obtain the injection current tensor as a function of photon energy. Convergence in $\v{k}$-points was achieved with a $70\times 70$ $\v{k}$-point mesh, see Fig.~\ref{fig:convergence}. In Fig.~\ref{fig:inj_curr_all}, we show $\eta_2^{yyz}$ of GeSe, GeS, SnS and SnSe monolayers. In all materials we see a very large response tensor (up to $10^{10}$ A/V$^2$s) that peaks in the visible light spectrum. In the Figure we also show the imaginary part of the dielectric function, which is a measure of the joint density of states and controls the optical absorption of the material. Close inspection reveals that only few peaks in $\varepsilon_2$ correspond to peaks in $\eta_2$ (as indicated by dashed lines) and in general, there is no obvious relation between the two.

\section{Electric polarization}
\label{sect:polarization_and_inj}
\begin{figure}[]
\includegraphics[width=0.45\textwidth]{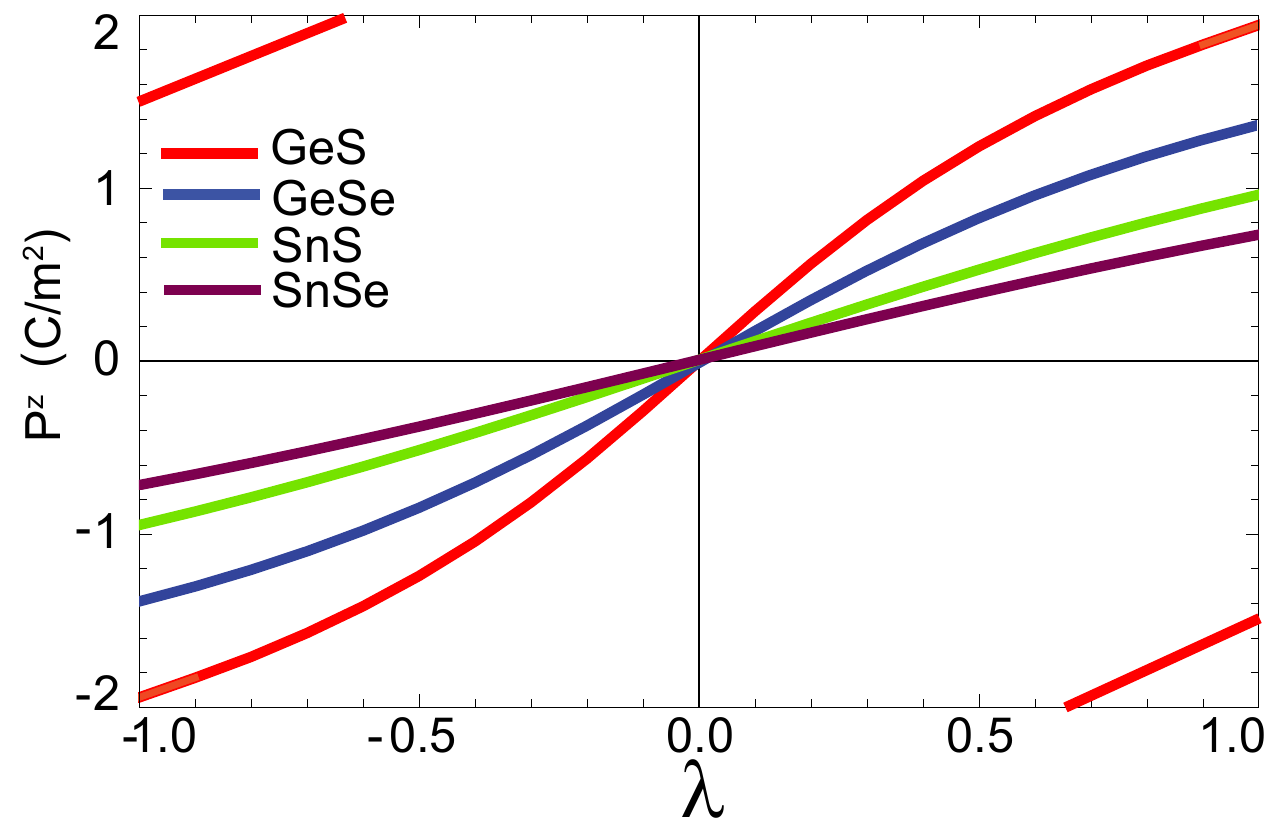}
\caption{Electric polarization along a path (parametrized by $\lambda$) between the asymmetric ground state $\lambda=\pm 1$ and the centrosymmetric configuration ($\lambda=0$) on the rectangular unit cell.}
\label{fig:pol_vs_lamda}
\end{figure}
The electric polarization of GeS, GeSe, SnS and SnSe is calculated numerically from Eqn.~\ref{eq:pol_formula}. The atomic positions in the centrosymmetric and noncentrosymmetric configurations are interpolated with straight lines under a constant area constraint~\cite{mm3}. Each atomic position is given by $\v{R}^i(\lambda)=\v{R}^i_0 + \lambda (\v{R}^i_f-\v{R}^i_0)$, where $\v{R}^i_0$ ($\v{R}^i_f$) is the initial (final) position of atom $i$ and $-1\leq \lambda\leq 1$ parametrizes the path in configuration space.

The use of straight lines to approximate the minimal energy path is expected to be a good approximation~\cite{Rangel2017,Wang2017}. Our spontaneous polarization values ($\lambda=1$) agrees with reports that follow a similar area constraint~\cite{Fei2016,Rangel2017}, see Table.~\ref{table:E_g_MMs}. The spontaneous polarization is found  from the difference in the polarization of the ground state with respect to that of the centrosymmetric configuration ($\lambda=0$), see Fig~\ref{fig:pol_vs_lamda}. For $\lambda<1$ the polarization is also calculated as the polarization difference with respect to the centrosymmetric configuration.

\LTcapwidth=0.45\textwidth
\begin{longtable}{lcccclcccclccc}
\caption{Positions (in {Bohr}) of atoms in single-layer monochalcogenides used to compute the polarization along a smooth path connecting $\v{R}^i_f(\lambda=-1)$ to $\v{R}^i_f(\lambda=1)$.}
\label{table:coords}
\endfirsthead
\multicolumn{9}{l}{\bf \textrm{GeS}}\\
\multicolumn{9}{l}{Lattice parameters:}\\
$\v{a} =$ & 28.34 & 0.00 & 0.00 \\
$\v{b} =$ &0.00   & 6.89 & 0.00 \\
$\v{c} =$ &0.00   & 0.00 & 8.52 \\
\multicolumn{9}{l}{Atom coordinates:}\\
&\multicolumn{3}{c}{$R_0$} &
\hspace{0.1in} &
\multicolumn{3}{c}{$R_f(\lambda=1.0)$}&
\hspace{0.1in} &
\multicolumn{3}{c}{$R_f(\lambda=-1.0)$}\\
\textrm{Ge} &
 2.66 & 1.72 & -0.01 &&
 2.66 & 1.72 &  1.14 &&
 2.66 & 1.72 & -1.16\\
\textrm{Ge} &
 7.50 & 5.17 & 4.25&&
 7.50 & 5.17 & 5.41&&
 7.50 & 5.17 & 3.09\\
\textrm{S}  &
 7.09 & 1.72 & -0.01&&
 7.09 & 1.72 & -0.01&&
 7.09 & 1.72 & -0.01\\
\textrm{S}  &
 3.06 & 5.17 & 4.25&&
 3.06 & 5.17 & 4.25&&
 3.06 & 5.17 & 4.25\\
\multicolumn{9}{l}{\bf \textrm{GeSe}}\\
$\v{a} =$ &28.83& 0.00 & 0.00 \\
$\v{b} =$ &0.00 & 7.50 & 0.00 \\
$\v{c} =$ &0.00 & 0.00 & 8.12 \\
\multicolumn{9}{l}{Atom coordinates:}\\
& \multicolumn{3}{c}{$R_0$} &
\hspace{0.1in} &
\multicolumn{3}{c}{$R_f(\lambda=1.0)$}&
\hspace{0.1in} &
\multicolumn{3}{c}{$R_f(\lambda=-1.0)$}\\
\textrm{Ge} &
 2.98 & 2.11 & 0.16 &&
 2.98 & 2.11 & 0.88 &&
 2.98 & 2.11 & -0.55\\
\textrm{Ge} &
 7.57 & 5.86 & 4.22&&
 7.57 & 5.86 & 4.94&&
 7.57 & 5.86 & 3.5\\
\textrm{Se} &
7.72 & 2.11 &  0.16&&
7.72 & 2.11 &  0.16&&
7.72 & 2.11 & 0.16\\
\textrm{Se} &
2.82 & 5.86 & 4.22&&
2.82 & 5.86 & 4.22&&
2.82 & 5.86 & 4.22\\
\multicolumn{9}{l}{\bf \textrm{SnS}}\\
\multicolumn{9}{l}{Lattice parameters:}\\
$\v{a} =$ &28.34& 0.00 & 0.00 \\
$\v{b} =$ &0.00 &  7.72 & 0.00 \\
$\v{c} =$ &0.00 & 0.00 & 8.12 \\
\multicolumn{9}{l}{Atom coordinates:}\\
& \multicolumn{3}{c}{$R_0$} &
\hspace{0.1in} &
\multicolumn{3}{c}{$R_f(\lambda=1.0)$}&
\hspace{0.1in} &
\multicolumn{3}{c}{$R_f(\lambda=-1.0)$}\\
\textrm{Sn} &
 2.73 & 1.93 & 0.31 &&
 2.73 & 1.93 & 0.94 &&
 2.73 & 1.93 & -0.30\\
\textrm{Sn} &
 8.12 & 5.79 & 4.37 &&
 8.12 & 5.79 & 5.00&&
 8.12 & 5.79 & 3.75\\
\textrm{S}  &
 7.60 & 1.93 & 0.31&&
 7.60 & 1.93 & 0.31&&
 7.60 & 1.93 & 0.31\\
\textrm{S}  &
 3.25 & 5.79 & 4.37&&
 3.25 & 5.79 & 4.37&&
 3.25 & 5.79 & 4.37\\
\multicolumn{9}{l}{\bf \textrm{SnSe}}\\
\multicolumn{9}{l}{Lattice parameters:}\\
$\v{a} =$ &28.34 & 0.00 & 0.00 \\
$\v{b} =$ &0.000 & 8.11 & 0.00 \\
$\v{c} =$ &0.000 & 0.00 & 8.31 \\
\multicolumn{9}{l}{Atom coordinates:}\\
& \multicolumn{3}{c}{$R_0$} &
\hspace{0.1in} &
\multicolumn{3}{c}{$R_f(\lambda=1.0)$}&
\hspace{0.1in} &
\multicolumn{3}{c}{$R_f(\lambda=-1.0)$}\\
\textrm{Sn} &
 2.98 & 2.11 & 0.29 &&
 2.98 & 2.11 & 0.71 &&
 2.98 & 2.11 & -0.13\\
\textrm{Sn} &
8.19 & 6.16 & 4.44 &&
8.19 & 6.16 & 4.87 &&
8.19 & 6.16 & 4.01\\
\textrm{Se} &
 8.12 & 2.11 & 0.29&&
 8.12 & 2.11 & 0.29&&
 8.12 & 2.11 & 0.29\\
\textrm{Se} &
 3.05 & 6.17 & 4.44&&
 3.05 & 6.17 & 4.44&&
 3.05 & 6.17 & 4.44\\
\end{longtable}

\section{Two-band model of the injection current of single-layer GeS}
\label{sec:tb_model}
We consider a two-band Hamiltonian

\begin{align}
H= f_0\sigma_0 + f_a \sigma_{a},
\end{align}
where $\sigma_a, a=x,y,z$ are the standard Pauli matrices and $\sigma_0$ is the $2\times 2$ identity matrix. Summation over repeated indices is implied. The functions $f_a$ are given by the  hopping integrals of the model. The Hamiltonian has eigenvectors given by
\begin{align}
u_c &= A
\begin{pmatrix}
f_x-if_y  \\
\epsilon - f_z
\end{pmatrix} \\
u_v &= A
\begin{pmatrix}
f_z-\epsilon  \\
f_x+if_y
\end{pmatrix},
\end{align}
where $A^{-2}=2\epsilon(\epsilon-f_z)$ is the normalization and $E_{c,v}=f_0 \pm \epsilon$ are the eivenvalues. $\epsilon=\sqrt{ f_a f_a}$ and $c,v$ denote the conduction and valence band respectively. An arbitrary phase factor has been omitted, since the final expression is independent of this phase. The Bloch wave functions are constructed as

\begin{align}
\psi_{n\v{k}} = \sum_{\v{R}} &e^{i\v{k}\cdot \v{R}} [ u_{n}^{(1)} \phi(\v{r}-\v{R}) \nn \\
&+ e^{i\v{k}\cdot \v{r}_0} u_{n}^{(2)} \phi(\v{r}-\v{r}_0-\v{R})],
\end{align}
where $u_{n}^{(i)}$ denotes the eigenvector corresponding to eigenvalue $n=v,c$ and $i=1,2$ denotes the first and second components. $\v{r}_0=(a_0,0)$ is the position of site $B$ with respect to site $A$ which is taken to be the origin. $\phi(\v{r})$ are $p_x$-orbitals and $\v{R}$ runs over all the A-type lattice positions. Notice that the phase of the wave function at site $B$ is different than that at site $A$~\cite{Bena2009}. 

Let us compute matrix elements in the expression for $\eta_2^{yyz}$ of the two-band model. Although this has been done before~\cite{Hosur2011,Juan2017}, here we present a very simple derivation. Denoting $X_{,a}\equiv \partial X/\partial k_a$ and using the definition of the Berry connection,  $r_{nm}^{a} = i \langle u_n| u_{m,a} \rangle$, we have
\begin{align}
r_{cv}^{z} r_{vc}^{y} - r_{cv}^{y} r_{vc}^{z} = 
- \langle u_c| u_{v,z}   \rangle  \langle u_v| u_{c,y}\rangle \nn \\
+ \langle u_c|u_{v,y} \rangle  \langle u_v|u_{c,z} \rangle. 
\end{align}
We can transfer derivatives from the ket to the bra at the expense of a minus as shown by taking derivatives of $\langle u_n| u_m \rangle=\delta_{nm}$. This, and the identity $1= |u_v \rangle \langle u_v| + |u_c \rangle \langle u_c|$ gives

\begin{align}
r_{cv}^{z} r_{vc}^{y} - r_{cv}^{y} r_{vc}^{z} &= 
\langle  u_{c,z}| u_{c,y} \rangle -\langle  u_{c,y}| u_{c,z} \rangle \nn\\
&\equiv i \Omega^{x}_{c},
\label{eq:rr_omega}
\end{align}
where in the last line we used the definition of the Berry curvature $\Omega_{n}^{a} = i \epsilon_{abc} \langle  u_{n,b}| u_{n,c} \rangle$. In general, one can show~\cite{Sipe2000}

\begin{align}
\Omega_{n}^{a} =i\epsilon_{abc} \sum_{m\neq n} r_{nm}^{b} r_{mn}^{c}, 
\end{align}
where $\epsilon_{abc}$ is the Levi-Civita tensor in three dimensions. Using ~\ref{eq:rr_omega} in Eq.~\ref{eq:eta2_yyz} we have, for the specific case of two bands,

\begin{align}
\eta_{2}^{yyz} = \frac{i e^3 \pi}{2 \hbar^2 V}\sum_{\v{k}} \omega_{cv,y}\Omega^{x}_{c}\delta(\omega_{cv}-\omega),
\label{eq:eta2_2B_v2}
\end{align} 
which is Eq.~\ref{eq:eta2_2B} of the main text. The Berry curvature in term of the $f$'s is $\Omega_{c}^{a} = \epsilon_{abc} \epsilon_{ajk} (A^2 f_i)_{,b}f_{k,c}$.

\begin{figure}
\includegraphics[width=.45\textwidth]{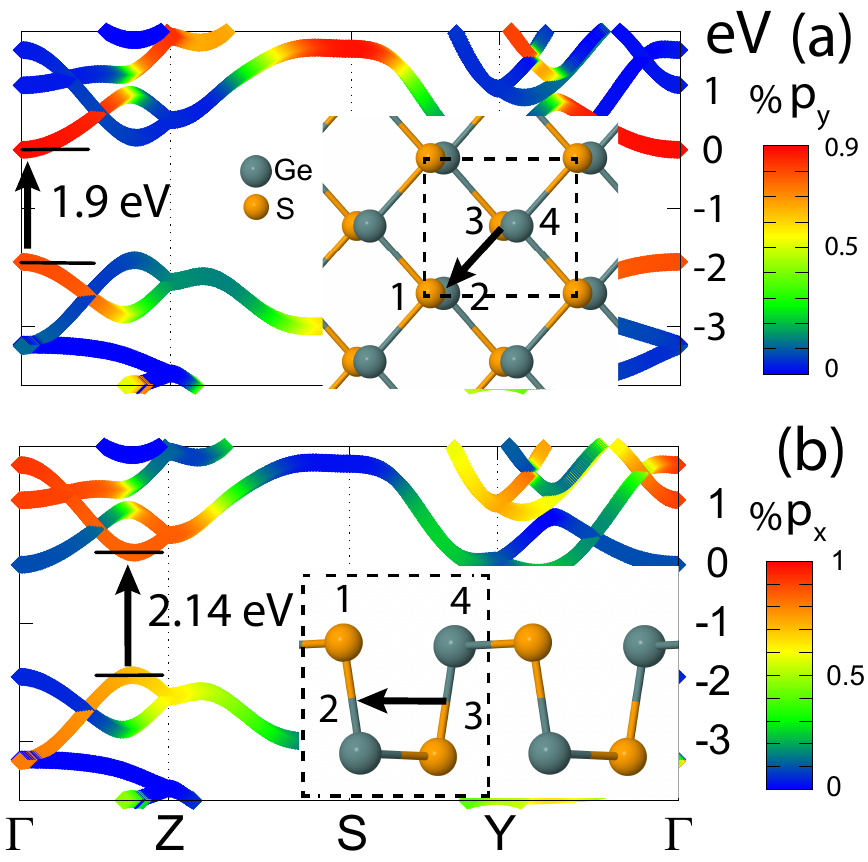}
\caption{Band projections into localized hydrogenic $s,p$ states for each of the Ge and S atoms in single-layer GeS. The states are centered at atomic positions 1-4 (inset). In (a) we show that the bands near the $\Gamma$ point in the BZ are mostly of $p_y$-character. However, the location of the valence and conduction band centers are at atomic positions 3,4 and 1,2 respectively. Hence the photoexcited carrier moves within the cell as indicated in the inset. In (b) we show that the bands near the Z point are mostly of $p_x$-character and that their valence and conduction band centers are also separated in real space. The transition at photon energies 2.14 eV gives a very large contribution to $\eta_2$ suggesting valence/conduction center separation plays an important role.}
\label{fig:ges_band_projection}
\end{figure}

\subsection{Parameters of single-layer GeS}
The effective Hamiltonian for monolayer GeS near the band edge is~\cite{Cook2017} 

\begin{align}
f_0 &= 2 t_{1}' (\cos\v{k}\cdot\v{a}_1 +\cos\v{k}\cdot\v{a}_2 )\nn \\
&~~~~~~~~+ 2 t_2' \cos\v{k}\cdot(\v{a}_1-\v{a}_2),\\
f_x - i f_y &= e^{i\v{k}\cdot \v{r}_0} (t_1  + t_2 \Phi_{\v{k}}+  t_3 \Phi_{\v{k}}^{*}), \\
f_z &= \Delta,
\end{align}
where $\Phi_{\v{k}}\equiv e^{-i\v{k}\cdot \v{a}_1} + e^{-i\v{k}\cdot \v{a}_2}$, $\Delta$ is the onsite potential and $t_1,t_2,t_3,t_1',t_2'$ are hopping matrix elements as indicated in Fig.~\ref{fig:tb_model}: $(t_1,t_2,t_3,t_1',t_2',\Delta)=(-2.33,0.61,0.13,0.07,-0.09,0.41)$ eV. $\v{a}_1 = (a_z,-a_y),\v{a}_2 = (a_z,a_y)$ are the primitive lattice vectors: $(a_z,a_y,d)=(4.53/2,3.63/2,2.56)$\AA~and $d$ is the thickness of the slab.  $\v{r}_0=(a_0,0)$ with $a_0=0.62$ \AA, is the relative position between sites A and site B. These parameters reproduce the DFT band structure and geometry of the wavefunction in the vicinity of the Gamma point~\cite{Cook2017}. The model is 2D; to compare with bulk values the results are multiplied by $2/d$ where the factor of 2 accounts for the smaller TB unit cell.

Note that the eigenvectors and eigenvalues are independent of $f_0$ and hence $t_{1}',t_{2}'$ do not enter into $\eta_2$. The parameter $a_0$ does not enter the band structure but it changes the wave function, e.g., it parametrizes the polarization of the system. Hence we can (loosely) associate $a_0$ with the contribution to $\eta_2$ from the polarization.

\subsection{Injection current near band edge of single-layer GeS}
An analytic expression for $\eta_2$ can be obtained for momenta near $\v{k}=0$. First let us expand $\omega_{cv}$ and the Berry curvature in small momenta  as

\begin{align}
\hbar\omega_{cv} &= E_g + \alpha a_z^2 k_z^2 + \beta a_y^2 k_y^2 + \gamma a_z^2 a_y^2 k_z^2  k_y^2 + ... \nn \\
\Omega^{x}_c &= (A a_0 + B a_z) a_y^2 k_y + C a_y^2 a_z^2 k_y  k_z^2 +...,
\end{align}
where $E_g= 1.89$ eV is the energy band gap,  ($\alpha,\beta,\gamma)=(2.30,1.33,-2.11)$ eV, $A=0.30, B=0.34$ are dimensionless constants that depend on the hopping parameters, and $C=-1.237 a_0 - 1.272 a_z$. Note that because of the mirror symmetry $y\to -y$ of the crystal the Berry curvature is odd under $k_y \to -k_y$ and $\omega_{cv}$ is even. Berry curvature is odd because in general $r_{nm}^{a}r_{mn}^{b}-r_{nm}^{b}r_{mn}^{a}$ is odd for $a=y$ or $b=y$ in the presence of this symmetry. To lowest order, the integrand of Eq.~\ref{eq:eta2_2B_v2} is quadratic in $k_y$ giving a leading linear $\omega - E_g$ contribution

\begin{align}
\eta_{2}^{yyz} &=  \frac{ie^3 \pi}{2 \hbar^2} \bigg[ \frac{a_y(A a_0 + B a_z)}{2\pi a_z d \sqrt{\alpha\beta}} (\hbar\omega-E_g) \nn \\  
&+ \frac{ a_y(\beta C + \gamma (A a_0 + B a_z)) }{8\pi a_z d \sqrt{\alpha \beta}\alpha\beta } (\hbar\omega-E_g)^2 +...\bigg],
\end{align}
for $\hbar \omega\geq E_g$. We have divided over $d/2$ to convert from 2D to bulk values. Substituting numerical values we obtain the result

\begin{align}
\eta_{2}^{yyz} &= \frac{ie^3 \pi}{2 \hbar^2} \bigg[ 0.05 \textrm{\AA} \left(\frac{\omega-E_g}{E_g}\right)\nn\\
&~~~~~~~~~~~~~~~-0.05 \textrm{\AA} \left(\frac{\omega-E_g}{E_g}\right)^2 +...\bigg].
\label{qe:eta2_analitical_tb}
\end{align}
which is plotted in Fig.~\ref{fig:tb_injection}. 


\begin{thebibliography}{61}%
\makeatletter
\providecommand \@ifxundefined [1]{%
 \@ifx{#1\undefined}
}%
\providecommand \@ifnum [1]{%
 \ifnum #1\expandafter \@firstoftwo
 \else \expandafter \@secondoftwo
 \fi
}%
\providecommand \@ifx [1]{%
 \ifx #1\expandafter \@firstoftwo
 \else \expandafter \@secondoftwo
 \fi
}%
\providecommand \natexlab [1]{#1}%
\providecommand \enquote  [1]{``#1''}%
\providecommand \bibnamefont  [1]{#1}%
\providecommand \bibfnamefont [1]{#1}%
\providecommand \citenamefont [1]{#1}%
\providecommand \href@noop [0]{\@secondoftwo}%
\providecommand \href [0]{\begingroup \@sanitize@url \@href}%
\providecommand \@href[1]{\@@startlink{#1}\@@href}%
\providecommand \@@href[1]{\endgroup#1\@@endlink}%
\providecommand \@sanitize@url [0]{\catcode `\\12\catcode `\$12\catcode
  `\&12\catcode `\#12\catcode `\^12\catcode `\_12\catcode `\%12\relax}%
\providecommand \@@startlink[1]{}%
\providecommand \@@endlink[0]{}%
\providecommand \url  [0]{\begingroup\@sanitize@url \@url }%
\providecommand \@url [1]{\endgroup\@href {#1}{\urlprefix }}%
\providecommand \urlprefix  [0]{URL }%
\providecommand \Eprint [0]{\href }%
\providecommand \doibase [0]{http://dx.doi.org/}%
\providecommand \selectlanguage [0]{\@gobble}%
\providecommand \bibinfo  [0]{\@secondoftwo}%
\providecommand \bibfield  [0]{\@secondoftwo}%
\providecommand \translation [1]{[#1]}%
\providecommand \BibitemOpen [0]{}%
\providecommand \bibitemStop [0]{}%
\providecommand \bibitemNoStop [0]{.\EOS\space}%
\providecommand \EOS [0]{\spacefactor3000\relax}%
\providecommand \BibitemShut  [1]{\csname bibitem#1\endcsname}%
\let\auto@bib@innerbib\@empty
\bibitem [{\citenamefont {Boyd}(2008)}]{Boyd2008}%
  \BibitemOpen
  \bibfield  {author} {\bibinfo {author} {\bibfnamefont {R.~W.}\ \bibnamefont
  {Boyd}},\ }\href@noop {} {\emph {\bibinfo {title} {Nonlinear Optics}}}\
  (\bibinfo  {publisher} {Academic Press},\ \bibinfo {year} {2008})\BibitemShut
  {NoStop}%
\bibitem [{\citenamefont {Sturman}\ and\ \citenamefont
  {Sturman}(1992)}]{Sturman1992}%
  \BibitemOpen
  \bibfield  {author} {\bibinfo {author} {\bibfnamefont {B.~I.}\ \bibnamefont
  {Sturman}}\ and\ \bibinfo {author} {\bibfnamefont {P.~J.}\ \bibnamefont
  {Sturman}},\ }\href@noop {} {\emph {\bibinfo {title} {Photovoltaic and
  Photo-refractive Effects in Noncentrosymmetric Materials}}}\ (\bibinfo
  {publisher} {CRC Press},\ \bibinfo {year} {1992})\BibitemShut {NoStop}%
\bibitem [{\citenamefont {Paillard}\ \emph {et~al.}(2018)\citenamefont
  {Paillard}, \citenamefont {Geneste}, \citenamefont {Bellaiche}, \citenamefont
  {Kreisel}, \citenamefont {Alexe},\ and\ \citenamefont
  {Dkhil}}]{Paillard2018}%
  \BibitemOpen
  \bibfield  {author} {\bibinfo {author} {\bibfnamefont {C.}~\bibnamefont
  {Paillard}}, \bibinfo {author} {\bibfnamefont {G.}~\bibnamefont {Geneste}},
  \bibinfo {author} {\bibfnamefont {L.}~\bibnamefont {Bellaiche}}, \bibinfo
  {author} {\bibfnamefont {J.}~\bibnamefont {Kreisel}}, \bibinfo {author}
  {\bibfnamefont {M.}~\bibnamefont {Alexe}}, \ and\ \bibinfo {author}
  {\bibfnamefont {B.}~\bibnamefont {Dkhil}},\ }\enquote {\bibinfo {title}
  {Emerging photovoltaic materials},}\ \ (\bibinfo  {publisher} {Wiley \& Sons
  Ltd},\ \bibinfo {year} {2018})\ Chap.~\bibinfo {chapter} {2}, p.\ \bibinfo
  {pages} {105}\BibitemShut {NoStop}%
\bibitem [{\citenamefont {Rioux}\ and\ \citenamefont {Sipe}(2012)}]{Rioux2012}%
  \BibitemOpen
  \bibfield  {author} {\bibinfo {author} {\bibfnamefont {J.}~\bibnamefont
  {Rioux}}\ and\ \bibinfo {author} {\bibfnamefont {J.}~\bibnamefont {Sipe}},\
  }\href@noop {} {\bibfield  {journal} {\bibinfo  {journal} {Physica E:
  Low-dimensional Systems and Nanostructures}\ }\textbf {\bibinfo {volume}
  {45}},\ \bibinfo {pages} {1} (\bibinfo {year} {2012})}\BibitemShut {NoStop}%
\bibitem [{\citenamefont {von Baltz}\ and\ \citenamefont
  {Kraut}(1981)}]{Baltz1981}%
  \BibitemOpen
  \bibfield  {author} {\bibinfo {author} {\bibfnamefont {R.}~\bibnamefont {von
  Baltz}}\ and\ \bibinfo {author} {\bibfnamefont {W.}~\bibnamefont {Kraut}},\
  }\href@noop {} {\bibfield  {journal} {\bibinfo  {journal} {Phys. Rev. B}\
  }\textbf {\bibinfo {volume} {23}},\ \bibinfo {pages} {5590} (\bibinfo {year}
  {1981})}\BibitemShut {NoStop}%
\bibitem [{\citenamefont {Sipe}\ and\ \citenamefont
  {Shkrebtii}(2000)}]{Sipe2000}%
  \BibitemOpen
  \bibfield  {author} {\bibinfo {author} {\bibfnamefont {J.~E.}\ \bibnamefont
  {Sipe}}\ and\ \bibinfo {author} {\bibfnamefont {A.~I.}\ \bibnamefont
  {Shkrebtii}},\ }\href@noop {} {\bibfield  {journal} {\bibinfo  {journal}
  {Phys. Rev. B}\ }\textbf {\bibinfo {volume} {61}},\ \bibinfo {pages} {5337}
  (\bibinfo {year} {2000})}\BibitemShut {NoStop}%
\bibitem [{\citenamefont {Morimoto}\ and\ \citenamefont
  {Nagaosa}(2016)}]{Morimoto2016}%
  \BibitemOpen
  \bibfield  {author} {\bibinfo {author} {\bibfnamefont {T.}~\bibnamefont
  {Morimoto}}\ and\ \bibinfo {author} {\bibfnamefont {N.}~\bibnamefont
  {Nagaosa}},\ }\href@noop {} {\bibfield  {journal} {\bibinfo  {journal}
  {Science Advances}\ }\textbf {\bibinfo {volume} {2}},\ \bibinfo {pages}
  {e1501524} (\bibinfo {year} {2016})}\BibitemShut {NoStop}%
\bibitem [{\citenamefont {Spanier}\ \emph {et~al.}(2016)\citenamefont
  {Spanier}, \citenamefont {Fridkin}, \citenamefont {Rappe}, \citenamefont
  {Akbashev}, \citenamefont {Polemi}, \citenamefont {Qi}, \citenamefont {Gu},
  \citenamefont {Young}, \citenamefont {Hawley}, \citenamefont {Imbrenda},
  \citenamefont {Xiao}, \citenamefont {Bennett-Jackson},\ and\ \citenamefont
  {Johnson}}]{Spanier2016}%
  \BibitemOpen
  \bibfield  {author} {\bibinfo {author} {\bibfnamefont {J.~E.}\ \bibnamefont
  {Spanier}}, \bibinfo {author} {\bibfnamefont {V.~M.}\ \bibnamefont
  {Fridkin}}, \bibinfo {author} {\bibfnamefont {A.~M.}\ \bibnamefont {Rappe}},
  \bibinfo {author} {\bibfnamefont {A.~R.}\ \bibnamefont {Akbashev}}, \bibinfo
  {author} {\bibfnamefont {A.}~\bibnamefont {Polemi}}, \bibinfo {author}
  {\bibfnamefont {Y.}~\bibnamefont {Qi}}, \bibinfo {author} {\bibfnamefont
  {Z.}~\bibnamefont {Gu}}, \bibinfo {author} {\bibfnamefont {S.~M.}\
  \bibnamefont {Young}}, \bibinfo {author} {\bibfnamefont {C.~J.}\ \bibnamefont
  {Hawley}}, \bibinfo {author} {\bibfnamefont {D.}~\bibnamefont {Imbrenda}},
  \bibinfo {author} {\bibfnamefont {G.}~\bibnamefont {Xiao}}, \bibinfo {author}
  {\bibfnamefont {A.~L.}\ \bibnamefont {Bennett-Jackson}}, \ and\ \bibinfo
  {author} {\bibfnamefont {C.~L.}\ \bibnamefont {Johnson}},\ }\href@noop {}
  {\bibfield  {journal} {\bibinfo  {journal} {Nat. Photonics}\ }\textbf
  {\bibinfo {volume} {10}},\ \bibinfo {pages} {611} (\bibinfo {year}
  {2016})}\BibitemShut {NoStop}%
\bibitem [{\citenamefont {Burger}\ \emph {et~al.}(2019)\citenamefont {Burger},
  \citenamefont {Agarwal}, \citenamefont {Aprelev}, \citenamefont {Schruba},
  \citenamefont {Gutierrez-Perez}, \citenamefont {Fridkin},\ and\ \citenamefont
  {Spanier}}]{Burger2019}%
  \BibitemOpen
  \bibfield  {author} {\bibinfo {author} {\bibfnamefont {A.~M.}\ \bibnamefont
  {Burger}}, \bibinfo {author} {\bibfnamefont {R.}~\bibnamefont {Agarwal}},
  \bibinfo {author} {\bibfnamefont {A.}~\bibnamefont {Aprelev}}, \bibinfo
  {author} {\bibfnamefont {E.}~\bibnamefont {Schruba}}, \bibinfo {author}
  {\bibfnamefont {A.}~\bibnamefont {Gutierrez-Perez}}, \bibinfo {author}
  {\bibfnamefont {V.~M.}\ \bibnamefont {Fridkin}}, \ and\ \bibinfo {author}
  {\bibfnamefont {J.~E.}\ \bibnamefont {Spanier}},\ }\href@noop {} {\bibfield
  {journal} {\bibinfo  {journal} {Science Advances}\ }\textbf {\bibinfo
  {volume} {5}},\ \bibinfo {pages} {eaau5588} (\bibinfo {year}
  {2019})}\BibitemShut {NoStop}%
\bibitem [{\citenamefont {Nakamura}\ \emph {et~al.}(2017)\citenamefont
  {Nakamura}, \citenamefont {Horiuchi}, \citenamefont {Kagawa}, \citenamefont
  {Ogawa}, \citenamefont {Kurumaji}, \citenamefont {Tokura},\ and\
  \citenamefont {Kawasaki}}]{Nakamura2017}%
  \BibitemOpen
  \bibfield  {author} {\bibinfo {author} {\bibfnamefont {M.}~\bibnamefont
  {Nakamura}}, \bibinfo {author} {\bibfnamefont {S.}~\bibnamefont {Horiuchi}},
  \bibinfo {author} {\bibfnamefont {F.}~\bibnamefont {Kagawa}}, \bibinfo
  {author} {\bibfnamefont {N.}~\bibnamefont {Ogawa}}, \bibinfo {author}
  {\bibfnamefont {T.}~\bibnamefont {Kurumaji}}, \bibinfo {author}
  {\bibfnamefont {Y.}~\bibnamefont {Tokura}}, \ and\ \bibinfo {author}
  {\bibfnamefont {M.}~\bibnamefont {Kawasaki}},\ }\href@noop {} {\bibfield
  {journal} {\bibinfo  {journal} {Nature Communications}\ }\textbf {\bibinfo
  {volume} {8}},\ \bibinfo {pages} {281} (\bibinfo {year} {2017})}\BibitemShut
  {NoStop}%
\bibitem [{\citenamefont {Ogawa}\ \emph {et~al.}(2017)\citenamefont {Ogawa},
  \citenamefont {Sotome}, \citenamefont {Kaneko}, \citenamefont {Ogino},\ and\
  \citenamefont {Tokura}}]{Ogawa2017}%
  \BibitemOpen
  \bibfield  {author} {\bibinfo {author} {\bibfnamefont {N.}~\bibnamefont
  {Ogawa}}, \bibinfo {author} {\bibfnamefont {M.}~\bibnamefont {Sotome}},
  \bibinfo {author} {\bibfnamefont {Y.}~\bibnamefont {Kaneko}}, \bibinfo
  {author} {\bibfnamefont {M.}~\bibnamefont {Ogino}}, \ and\ \bibinfo {author}
  {\bibfnamefont {Y.}~\bibnamefont {Tokura}},\ }\href@noop {} {\bibfield
  {journal} {\bibinfo  {journal} {Phys. Rev. B}\ }\textbf {\bibinfo {volume}
  {96}},\ \bibinfo {pages} {241203} (\bibinfo {year} {2017})}\BibitemShut
  {NoStop}%
\bibitem [{\citenamefont {Nagaosa}\ and\ \citenamefont
  {Morimoto}(2017)}]{Nagaosa2017}%
  \BibitemOpen
  \bibfield  {author} {\bibinfo {author} {\bibfnamefont {N.}~\bibnamefont
  {Nagaosa}}\ and\ \bibinfo {author} {\bibfnamefont {T.}~\bibnamefont
  {Morimoto}},\ }\href@noop {} {\bibfield  {journal} {\bibinfo  {journal}
  {Advanced Materials}\ }\textbf {\bibinfo {volume} {29}},\ \bibinfo {pages}
  {1603345} (\bibinfo {year} {2017})}\BibitemShut {NoStop}%
\bibitem [{\citenamefont {Fregoso}\ \emph {et~al.}(2017)\citenamefont
  {Fregoso}, \citenamefont {Morimoto},\ and\ \citenamefont
  {Moore}}]{Fregoso2017}%
  \BibitemOpen
  \bibfield  {author} {\bibinfo {author} {\bibfnamefont {B.~M.}\ \bibnamefont
  {Fregoso}}, \bibinfo {author} {\bibfnamefont {T.}~\bibnamefont {Morimoto}}, \
  and\ \bibinfo {author} {\bibfnamefont {J.~E.}\ \bibnamefont {Moore}},\
  }\href@noop {} {\bibfield  {journal} {\bibinfo  {journal} {Phys. Rev. B}\
  }\textbf {\bibinfo {volume} {96}},\ \bibinfo {pages} {075421} (\bibinfo
  {year} {2017})}\BibitemShut {NoStop}%
\bibitem [{\citenamefont {Rangel}\ \emph {et~al.}(2017)\citenamefont {Rangel},
  \citenamefont {Fregoso}, \citenamefont {Mendoza}, \citenamefont {Morimoto},
  \citenamefont {Moore},\ and\ \citenamefont {Neaton}}]{Rangel2017}%
  \BibitemOpen
  \bibfield  {author} {\bibinfo {author} {\bibfnamefont {T.}~\bibnamefont
  {Rangel}}, \bibinfo {author} {\bibfnamefont {B.~M.}\ \bibnamefont {Fregoso}},
  \bibinfo {author} {\bibfnamefont {B.~S.}\ \bibnamefont {Mendoza}}, \bibinfo
  {author} {\bibfnamefont {T.}~\bibnamefont {Morimoto}}, \bibinfo {author}
  {\bibfnamefont {J.~E.}\ \bibnamefont {Moore}}, \ and\ \bibinfo {author}
  {\bibfnamefont {J.~B.}\ \bibnamefont {Neaton}},\ }\href@noop {} {\bibfield
  {journal} {\bibinfo  {journal} {Phys. Rev. Lett.}\ }\textbf {\bibinfo
  {volume} {119}},\ \bibinfo {pages} {067402} (\bibinfo {year}
  {2017})}\BibitemShut {NoStop}%
\bibitem [{\citenamefont {Wang}\ \emph {et~al.}(2017)\citenamefont {Wang},
  \citenamefont {Liu}, \citenamefont {Kang}, \citenamefont {Gu}, \citenamefont
  {Xu},\ and\ \citenamefont {Duan}}]{Wang2017b}%
  \BibitemOpen
  \bibfield  {author} {\bibinfo {author} {\bibfnamefont {C.}~\bibnamefont
  {Wang}}, \bibinfo {author} {\bibfnamefont {X.}~\bibnamefont {Liu}}, \bibinfo
  {author} {\bibfnamefont {L.}~\bibnamefont {Kang}}, \bibinfo {author}
  {\bibfnamefont {B.-L.}\ \bibnamefont {Gu}}, \bibinfo {author} {\bibfnamefont
  {Y.}~\bibnamefont {Xu}}, \ and\ \bibinfo {author} {\bibfnamefont
  {W.}~\bibnamefont {Duan}},\ }\href@noop {} {\bibfield  {journal} {\bibinfo
  {journal} {Phys. Rev. B}\ }\textbf {\bibinfo {volume} {96}},\ \bibinfo
  {pages} {115147} (\bibinfo {year} {2017})}\BibitemShut {NoStop}%
\bibitem [{\citenamefont {Kushnir}\ \emph {et~al.}(2017)\citenamefont
  {Kushnir}, \citenamefont {Wang}, \citenamefont {Fitzgerald}, \citenamefont
  {Koski},\ and\ \citenamefont {Titova}}]{Kushnir2017}%
  \BibitemOpen
  \bibfield  {author} {\bibinfo {author} {\bibfnamefont {K.}~\bibnamefont
  {Kushnir}}, \bibinfo {author} {\bibfnamefont {M.}~\bibnamefont {Wang}},
  \bibinfo {author} {\bibfnamefont {P.~D.}\ \bibnamefont {Fitzgerald}},
  \bibinfo {author} {\bibfnamefont {K.~J.}\ \bibnamefont {Koski}}, \ and\
  \bibinfo {author} {\bibfnamefont {L.~V.}\ \bibnamefont {Titova}},\
  }\href@noop {} {\bibfield  {journal} {\bibinfo  {journal} {ACS Energy
  Letters}\ }\textbf {\bibinfo {volume} {2}},\ \bibinfo {pages} {1429}
  (\bibinfo {year} {2017})}\BibitemShut {NoStop}%
\bibitem [{\citenamefont {Nakamura}\ \emph {et~al.}(2018)\citenamefont
  {Nakamura}, \citenamefont {Hatada}, \citenamefont {Kaneko}, \citenamefont
  {Ogawa}, \citenamefont {Tokura},\ and\ \citenamefont
  {Kawasaki}}]{Nakamura2018}%
  \BibitemOpen
  \bibfield  {author} {\bibinfo {author} {\bibfnamefont {M.}~\bibnamefont
  {Nakamura}}, \bibinfo {author} {\bibfnamefont {H.}~\bibnamefont {Hatada}},
  \bibinfo {author} {\bibfnamefont {Y.}~\bibnamefont {Kaneko}}, \bibinfo
  {author} {\bibfnamefont {N.}~\bibnamefont {Ogawa}}, \bibinfo {author}
  {\bibfnamefont {Y.}~\bibnamefont {Tokura}}, \ and\ \bibinfo {author}
  {\bibfnamefont {M.}~\bibnamefont {Kawasaki}},\ }\href@noop {} {\bibfield
  {journal} {\bibinfo  {journal} {Applied Physics Letters}\ }\textbf {\bibinfo
  {volume} {113}},\ \bibinfo {pages} {232901} (\bibinfo {year}
  {2018})}\BibitemShut {NoStop}%
\bibitem [{\citenamefont {Gong}\ \emph {et~al.}(2018)\citenamefont {Gong},
  \citenamefont {Zheng},\ and\ \citenamefont {Rappe}}]{Gong2018}%
  \BibitemOpen
  \bibfield  {author} {\bibinfo {author} {\bibfnamefont {S.-J.}\ \bibnamefont
  {Gong}}, \bibinfo {author} {\bibfnamefont {F.}~\bibnamefont {Zheng}}, \ and\
  \bibinfo {author} {\bibfnamefont {A.~M.}\ \bibnamefont {Rappe}},\ }\href@noop
  {} {\bibfield  {journal} {\bibinfo  {journal} {Phys. Rev. Lett.}\ }\textbf
  {\bibinfo {volume} {121}},\ \bibinfo {pages} {017402} (\bibinfo {year}
  {2018})}\BibitemShut {NoStop}%
\bibitem [{\citenamefont {Kushnir}\ \emph {et~al.}(2019)\citenamefont
  {Kushnir}, \citenamefont {Qin}, \citenamefont {Shen}, \citenamefont {Li},
  \citenamefont {Fregoso}, \citenamefont {Tongay},\ and\ \citenamefont
  {Titova}}]{Kushnir2019}%
  \BibitemOpen
  \bibfield  {author} {\bibinfo {author} {\bibfnamefont {K.}~\bibnamefont
  {Kushnir}}, \bibinfo {author} {\bibfnamefont {Y.}~\bibnamefont {Qin}},
  \bibinfo {author} {\bibfnamefont {Y.}~\bibnamefont {Shen}}, \bibinfo {author}
  {\bibfnamefont {G.}~\bibnamefont {Li}}, \bibinfo {author} {\bibfnamefont
  {B.~M.}\ \bibnamefont {Fregoso}}, \bibinfo {author} {\bibfnamefont
  {S.}~\bibnamefont {Tongay}}, \ and\ \bibinfo {author} {\bibfnamefont {L.~V.}\
  \bibnamefont {Titova}},\ }\href@noop {} {\bibfield  {journal} {\bibinfo
  {journal} {ACS Applied Materials \& Interfaces}\ }\textbf {\bibinfo {volume}
  {11}},\ \bibinfo {pages} {5492} (\bibinfo {year} {2019})}\BibitemShut
  {NoStop}%
\bibitem [{\citenamefont {Sotome}\ \emph
  {et~al.}(2019{\natexlab{a}})\citenamefont {Sotome}, \citenamefont {Nakamura},
  \citenamefont {Fujioka}, \citenamefont {Ogino}, \citenamefont {Kaneko},
  \citenamefont {Morimoto}, \citenamefont {Zhang}, \citenamefont {Kawasaki},
  \citenamefont {Nagaosa}, \citenamefont {Tokura},\ and\ \citenamefont
  {Ogawa}}]{Sotome2019}%
  \BibitemOpen
  \bibfield  {author} {\bibinfo {author} {\bibfnamefont {M.}~\bibnamefont
  {Sotome}}, \bibinfo {author} {\bibfnamefont {M.}~\bibnamefont {Nakamura}},
  \bibinfo {author} {\bibfnamefont {J.}~\bibnamefont {Fujioka}}, \bibinfo
  {author} {\bibfnamefont {M.}~\bibnamefont {Ogino}}, \bibinfo {author}
  {\bibfnamefont {Y.}~\bibnamefont {Kaneko}}, \bibinfo {author} {\bibfnamefont
  {T.}~\bibnamefont {Morimoto}}, \bibinfo {author} {\bibfnamefont
  {Y.}~\bibnamefont {Zhang}}, \bibinfo {author} {\bibfnamefont
  {M.}~\bibnamefont {Kawasaki}}, \bibinfo {author} {\bibfnamefont
  {N.}~\bibnamefont {Nagaosa}}, \bibinfo {author} {\bibfnamefont
  {Y.}~\bibnamefont {Tokura}}, \ and\ \bibinfo {author} {\bibfnamefont
  {N.}~\bibnamefont {Ogawa}},\ }\href@noop {} {\bibfield  {journal} {\bibinfo
  {journal} {Proceedings of the National Academy of Sciences}\ }\textbf
  {\bibinfo {volume} {116}},\ \bibinfo {pages} {1929} (\bibinfo {year}
  {2019}{\natexlab{a}})}\BibitemShut {NoStop}%
\bibitem [{\citenamefont {Sotome}\ \emph
  {et~al.}(2019{\natexlab{b}})\citenamefont {Sotome}, \citenamefont {Nakamura},
  \citenamefont {Fujioka}, \citenamefont {Ogino}, \citenamefont {Kaneko},
  \citenamefont {Morimoto}, \citenamefont {Zhang}, \citenamefont {Kawasaki},
  \citenamefont {Nagaosa}, \citenamefont {Tokura},\ and\ \citenamefont
  {Ogawa}}]{Sotome2019a}%
  \BibitemOpen
  \bibfield  {author} {\bibinfo {author} {\bibfnamefont {M.}~\bibnamefont
  {Sotome}}, \bibinfo {author} {\bibfnamefont {M.}~\bibnamefont {Nakamura}},
  \bibinfo {author} {\bibfnamefont {J.}~\bibnamefont {Fujioka}}, \bibinfo
  {author} {\bibfnamefont {M.}~\bibnamefont {Ogino}}, \bibinfo {author}
  {\bibfnamefont {Y.}~\bibnamefont {Kaneko}}, \bibinfo {author} {\bibfnamefont
  {T.}~\bibnamefont {Morimoto}}, \bibinfo {author} {\bibfnamefont
  {Y.}~\bibnamefont {Zhang}}, \bibinfo {author} {\bibfnamefont
  {M.}~\bibnamefont {Kawasaki}}, \bibinfo {author} {\bibfnamefont
  {N.}~\bibnamefont {Nagaosa}}, \bibinfo {author} {\bibfnamefont
  {Y.}~\bibnamefont {Tokura}}, \ and\ \bibinfo {author} {\bibfnamefont
  {N.}~\bibnamefont {Ogawa}},\ }\href@noop {} {\bibfield  {journal} {\bibinfo
  {journal} {Applied Physics Letters}\ }\textbf {\bibinfo {volume} {114}},\
  \bibinfo {pages} {151101} (\bibinfo {year} {2019}{\natexlab{b}})}\BibitemShut
  {NoStop}%
\bibitem [{\citenamefont {Kr\'al}\ \emph {et~al.}(2000)\citenamefont {Kr\'al},
  \citenamefont {Mele},\ and\ \citenamefont {Tom\'anek}}]{Kral2000}%
  \BibitemOpen
  \bibfield  {author} {\bibinfo {author} {\bibfnamefont {P.}~\bibnamefont
  {Kr\'al}}, \bibinfo {author} {\bibfnamefont {E.~J.}\ \bibnamefont {Mele}}, \
  and\ \bibinfo {author} {\bibfnamefont {D.}~\bibnamefont {Tom\'anek}},\
  }\href@noop {} {\bibfield  {journal} {\bibinfo  {journal} {Phys. Rev. Lett.}\
  }\textbf {\bibinfo {volume} {85}},\ \bibinfo {pages} {1512} (\bibinfo {year}
  {2000})}\BibitemShut {NoStop}%
\bibitem [{\citenamefont {Cook}\ \emph {et~al.}(2017)\citenamefont {Cook},
  \citenamefont {M.~Fregoso}, \citenamefont {de~Juan}, \citenamefont {Coh},\
  and\ \citenamefont {Moore}}]{Cook2017}%
  \BibitemOpen
  \bibfield  {author} {\bibinfo {author} {\bibfnamefont {A.~M.}\ \bibnamefont
  {Cook}}, \bibinfo {author} {\bibfnamefont {B.}~\bibnamefont {M.~Fregoso}},
  \bibinfo {author} {\bibfnamefont {F.}~\bibnamefont {de~Juan}}, \bibinfo
  {author} {\bibfnamefont {S.}~\bibnamefont {Coh}}, \ and\ \bibinfo {author}
  {\bibfnamefont {J.~E.}\ \bibnamefont {Moore}},\ }\href@noop {} {\bibfield
  {journal} {\bibinfo  {journal} {Nature Communications}\ }\textbf {\bibinfo
  {volume} {8}},\ \bibinfo {pages} {14176} (\bibinfo {year}
  {2017})}\BibitemShut {NoStop}%
\bibitem [{\citenamefont {Zhang}\ \emph {et~al.}(2019)\citenamefont {Zhang},
  \citenamefont {Ideue}, \citenamefont {Onga}, \citenamefont {Qin},
  \citenamefont {Suzuki}, \citenamefont {Zak}, \citenamefont {Tenne},
  \citenamefont {Smet},\ and\ \citenamefont {Iwasa}}]{Zhang2019}%
  \BibitemOpen
  \bibfield  {author} {\bibinfo {author} {\bibfnamefont {Y.~J.}\ \bibnamefont
  {Zhang}}, \bibinfo {author} {\bibfnamefont {T.}~\bibnamefont {Ideue}},
  \bibinfo {author} {\bibfnamefont {M.}~\bibnamefont {Onga}}, \bibinfo {author}
  {\bibfnamefont {F.}~\bibnamefont {Qin}}, \bibinfo {author} {\bibfnamefont
  {R.}~\bibnamefont {Suzuki}}, \bibinfo {author} {\bibfnamefont
  {A.}~\bibnamefont {Zak}}, \bibinfo {author} {\bibfnamefont {R.}~\bibnamefont
  {Tenne}}, \bibinfo {author} {\bibfnamefont {J.~H.}\ \bibnamefont {Smet}}, \
  and\ \bibinfo {author} {\bibfnamefont {Y.}~\bibnamefont {Iwasa}},\
  }\href@noop {} {\bibfield  {journal} {\bibinfo  {journal} {Nature}\ }\textbf
  {\bibinfo {volume} {570}},\ \bibinfo {pages} {349} (\bibinfo {year}
  {2019})}\BibitemShut {NoStop}%
\bibitem [{\citenamefont {Iba\~nez Azpiroz}\ \emph {et~al.}(2018)\citenamefont
  {Iba\~nez Azpiroz}, \citenamefont {Tsirkin},\ and\ \citenamefont
  {Souza}}]{Ibanez-Azpiroz2018}%
  \BibitemOpen
  \bibfield  {author} {\bibinfo {author} {\bibfnamefont {J.}~\bibnamefont
  {Iba\~nez Azpiroz}}, \bibinfo {author} {\bibfnamefont {S.~S.}\ \bibnamefont
  {Tsirkin}}, \ and\ \bibinfo {author} {\bibfnamefont {I.}~\bibnamefont
  {Souza}},\ }\href@noop {} {\bibfield  {journal} {\bibinfo  {journal} {Phys.
  Rev. B}\ }\textbf {\bibinfo {volume} {97}},\ \bibinfo {pages} {245143}
  (\bibinfo {year} {2018})}\BibitemShut {NoStop}%
\bibitem [{\citenamefont {Brehm}(2018)}]{Brehm2018}%
  \BibitemOpen
  \bibfield  {author} {\bibinfo {author} {\bibfnamefont {J.~A.}\ \bibnamefont
  {Brehm}},\ }\href@noop {} {\bibfield  {journal} {\bibinfo  {journal} {J.
  Mater. Chem. C}\ }\textbf {\bibinfo {volume} {6}},\ \bibinfo {pages} {1470}
  (\bibinfo {year} {2018})}\BibitemShut {NoStop}%
\bibitem [{\citenamefont {Hosur}(2011)}]{Hosur2011}%
  \BibitemOpen
  \bibfield  {author} {\bibinfo {author} {\bibfnamefont {P.}~\bibnamefont
  {Hosur}},\ }\href@noop {} {\bibfield  {journal} {\bibinfo  {journal} {Phys.
  Rev. B}\ }\textbf {\bibinfo {volume} {83}},\ \bibinfo {pages} {035309}
  (\bibinfo {year} {2011})}\BibitemShut {NoStop}%
\bibitem [{\citenamefont {de~Juan}\ \emph {et~al.}(2017)\citenamefont
  {de~Juan}, \citenamefont {Grushin}, \citenamefont {Morimoto},\ and\
  \citenamefont {Moore}}]{Juan2017}%
  \BibitemOpen
  \bibfield  {author} {\bibinfo {author} {\bibfnamefont {F.}~\bibnamefont
  {de~Juan}}, \bibinfo {author} {\bibfnamefont {A.~G.}\ \bibnamefont
  {Grushin}}, \bibinfo {author} {\bibfnamefont {T.}~\bibnamefont {Morimoto}}, \
  and\ \bibinfo {author} {\bibfnamefont {J.~E.}\ \bibnamefont {Moore}},\
  }\href@noop {} {\bibfield  {journal} {\bibinfo  {journal} {Nature
  Communications}\ }\textbf {\bibinfo {volume} {8}},\ \bibinfo {pages} {15995}
  (\bibinfo {year} {2017})}\BibitemShut {NoStop}%
\bibitem [{\citenamefont {Rees}\ \emph {et~al.}()\citenamefont {Rees},
  \citenamefont {Manna}, \citenamefont {Lu}, \citenamefont {Morimoto},
  \citenamefont {Borrmann}, \citenamefont {Felser}, \citenamefont {Moore},
  \citenamefont {Torchinsky},\ and\ \citenamefont {Orenstein}}]{Rees}%
  \BibitemOpen
  \bibfield  {author} {\bibinfo {author} {\bibfnamefont {D.}~\bibnamefont
  {Rees}}, \bibinfo {author} {\bibfnamefont {K.}~\bibnamefont {Manna}},
  \bibinfo {author} {\bibfnamefont {B.}~\bibnamefont {Lu}}, \bibinfo {author}
  {\bibfnamefont {T.}~\bibnamefont {Morimoto}}, \bibinfo {author}
  {\bibfnamefont {H.}~\bibnamefont {Borrmann}}, \bibinfo {author}
  {\bibfnamefont {C.}~\bibnamefont {Felser}}, \bibinfo {author} {\bibfnamefont
  {J.}~\bibnamefont {Moore}}, \bibinfo {author} {\bibfnamefont {D.~H.}\
  \bibnamefont {Torchinsky}}, \ and\ \bibinfo {author} {\bibfnamefont
  {J.}~\bibnamefont {Orenstein}},\ }\href@noop {} {}\bibinfo {note}
  {ArXiv:1902.03230 [cond-mat.mes-hall]}\BibitemShut {NoStop}%
\bibitem [{\citenamefont {Chan}\ \emph {et~al.}(2017)\citenamefont {Chan},
  \citenamefont {Lindner}, \citenamefont {Refael},\ and\ \citenamefont
  {Lee}}]{Chan2017}%
  \BibitemOpen
  \bibfield  {author} {\bibinfo {author} {\bibfnamefont {C.-K.}\ \bibnamefont
  {Chan}}, \bibinfo {author} {\bibfnamefont {N.~H.}\ \bibnamefont {Lindner}},
  \bibinfo {author} {\bibfnamefont {G.}~\bibnamefont {Refael}}, \ and\ \bibinfo
  {author} {\bibfnamefont {P.~A.}\ \bibnamefont {Lee}},\ }\href@noop {}
  {\bibfield  {journal} {\bibinfo  {journal} {Phys. Rev. B}\ }\textbf {\bibinfo
  {volume} {95}},\ \bibinfo {pages} {041104} (\bibinfo {year}
  {2017})}\BibitemShut {NoStop}%
\bibitem [{\citenamefont {Flicker}\ \emph {et~al.}(2018)\citenamefont
  {Flicker}, \citenamefont {de~Juan}, \citenamefont {Bradlyn}, \citenamefont
  {Morimoto}, \citenamefont {Vergniory},\ and\ \citenamefont
  {Grushin}}]{Flicker2018}%
  \BibitemOpen
  \bibfield  {author} {\bibinfo {author} {\bibfnamefont {F.}~\bibnamefont
  {Flicker}}, \bibinfo {author} {\bibfnamefont {F.}~\bibnamefont {de~Juan}},
  \bibinfo {author} {\bibfnamefont {B.}~\bibnamefont {Bradlyn}}, \bibinfo
  {author} {\bibfnamefont {T.}~\bibnamefont {Morimoto}}, \bibinfo {author}
  {\bibfnamefont {M.~G.}\ \bibnamefont {Vergniory}}, \ and\ \bibinfo {author}
  {\bibfnamefont {A.~G.}\ \bibnamefont {Grushin}},\ }\href@noop {} {\bibfield
  {journal} {\bibinfo  {journal} {Phys. Rev. B}\ }\textbf {\bibinfo {volume}
  {98}},\ \bibinfo {pages} {155145} (\bibinfo {year} {2018})}\BibitemShut
  {NoStop}%
\bibitem [{\citenamefont {Parker}\ \emph {et~al.}(2019)\citenamefont {Parker},
  \citenamefont {Morimoto}, \citenamefont {Orenstein},\ and\ \citenamefont
  {Moore}}]{Parker2019}%
  \BibitemOpen
  \bibfield  {author} {\bibinfo {author} {\bibfnamefont {D.~E.}\ \bibnamefont
  {Parker}}, \bibinfo {author} {\bibfnamefont {T.}~\bibnamefont {Morimoto}},
  \bibinfo {author} {\bibfnamefont {J.}~\bibnamefont {Orenstein}}, \ and\
  \bibinfo {author} {\bibfnamefont {J.~E.}\ \bibnamefont {Moore}},\ }\href@noop
  {} {\bibfield  {journal} {\bibinfo  {journal} {Phys. Rev. B}\ }\textbf
  {\bibinfo {volume} {99}},\ \bibinfo {pages} {045121} (\bibinfo {year}
  {2019})}\BibitemShut {NoStop}%
\bibitem [{\citenamefont {Tan}\ and\ \citenamefont {Rappe}(2019)}]{Tan2019}%
  \BibitemOpen
  \bibfield  {author} {\bibinfo {author} {\bibfnamefont {L.~Z.}\ \bibnamefont
  {Tan}}\ and\ \bibinfo {author} {\bibfnamefont {A.~M.}\ \bibnamefont
  {Rappe}},\ }\href@noop {} {\bibfield  {journal} {\bibinfo  {journal} {Journal
  of Physics: Condensed Matter}\ }\textbf {\bibinfo {volume} {31}},\ \bibinfo
  {pages} {084002} (\bibinfo {year} {2019})}\BibitemShut {NoStop}%
\bibitem [{\citenamefont {Barik}\ and\ \citenamefont {Sau}()}]{Barik}%
  \BibitemOpen
  \bibfield  {author} {\bibinfo {author} {\bibfnamefont {T.}~\bibnamefont
  {Barik}}\ and\ \bibinfo {author} {\bibfnamefont {J.~D.}\ \bibnamefont
  {Sau}},\ }\href@noop {} {}\bibinfo {note} {ArXiv:1908.01793
  [cond-mat.mes-hall]}\BibitemShut {NoStop}%
\bibitem [{\citenamefont {Chang}\ \emph {et~al.}(2016)\citenamefont {Chang},
  \citenamefont {Liu}, \citenamefont {Lin}, \citenamefont {Wang}, \citenamefont
  {Zhao}, \citenamefont {Zhang}, \citenamefont {Jin}, \citenamefont {Zhong},
  \citenamefont {Hu}, \citenamefont {Duan}, \citenamefont {Zhang},
  \citenamefont {Fu}, \citenamefont {Xue}, \citenamefont {Chen},\ and\
  \citenamefont {Ji}}]{Chang2016}%
  \BibitemOpen
  \bibfield  {author} {\bibinfo {author} {\bibfnamefont {K.}~\bibnamefont
  {Chang}}, \bibinfo {author} {\bibfnamefont {J.}~\bibnamefont {Liu}}, \bibinfo
  {author} {\bibfnamefont {H.}~\bibnamefont {Lin}}, \bibinfo {author}
  {\bibfnamefont {N.}~\bibnamefont {Wang}}, \bibinfo {author} {\bibfnamefont
  {K.}~\bibnamefont {Zhao}}, \bibinfo {author} {\bibfnamefont {A.}~\bibnamefont
  {Zhang}}, \bibinfo {author} {\bibfnamefont {F.}~\bibnamefont {Jin}}, \bibinfo
  {author} {\bibfnamefont {Y.}~\bibnamefont {Zhong}}, \bibinfo {author}
  {\bibfnamefont {X.}~\bibnamefont {Hu}}, \bibinfo {author} {\bibfnamefont
  {W.}~\bibnamefont {Duan}}, \bibinfo {author} {\bibfnamefont {Q.}~\bibnamefont
  {Zhang}}, \bibinfo {author} {\bibfnamefont {L.}~\bibnamefont {Fu}}, \bibinfo
  {author} {\bibfnamefont {Q.-K.}\ \bibnamefont {Xue}}, \bibinfo {author}
  {\bibfnamefont {X.}~\bibnamefont {Chen}}, \ and\ \bibinfo {author}
  {\bibfnamefont {S.-H.}\ \bibnamefont {Ji}},\ }\href@noop {} {\bibfield
  {journal} {\bibinfo  {journal} {Science}\ }\textbf {\bibinfo {volume}
  {353}},\ \bibinfo {pages} {274} (\bibinfo {year} {2016})}\BibitemShut
  {NoStop}%
\bibitem [{\citenamefont {Mehboudi}\ \emph {et~al.}(2016)\citenamefont
  {Mehboudi}, \citenamefont {Fregoso}, \citenamefont {Yang}, \citenamefont
  {Zhu}, \citenamefont {van~der Zande}, \citenamefont {Ferrer}, \citenamefont
  {Bellaiche}, \citenamefont {Kumar},\ and\ \citenamefont
  {Barraza-Lopez}}]{mm2}%
  \BibitemOpen
  \bibfield  {author} {\bibinfo {author} {\bibfnamefont {M.}~\bibnamefont
  {Mehboudi}}, \bibinfo {author} {\bibfnamefont {B.~M.}\ \bibnamefont
  {Fregoso}}, \bibinfo {author} {\bibfnamefont {Y.}~\bibnamefont {Yang}},
  \bibinfo {author} {\bibfnamefont {W.}~\bibnamefont {Zhu}}, \bibinfo {author}
  {\bibfnamefont {A.}~\bibnamefont {van~der Zande}}, \bibinfo {author}
  {\bibfnamefont {J.}~\bibnamefont {Ferrer}}, \bibinfo {author} {\bibfnamefont
  {L.}~\bibnamefont {Bellaiche}}, \bibinfo {author} {\bibfnamefont
  {P.}~\bibnamefont {Kumar}}, \ and\ \bibinfo {author} {\bibfnamefont
  {S.}~\bibnamefont {Barraza-Lopez}},\ }\href@noop {} {\bibfield  {journal}
  {\bibinfo  {journal} {Phys. Rev. Lett.}\ }\textbf {\bibinfo {volume} {117}},\
  \bibinfo {pages} {246802} (\bibinfo {year} {2016})}\BibitemShut {NoStop}%
\bibitem [{\citenamefont {Barraza-Lopez}\ \emph {et~al.}(2018)\citenamefont
  {Barraza-Lopez}, \citenamefont {Kaloni}, \citenamefont {Poudel},\ and\
  \citenamefont {Kumar}}]{mm3}%
  \BibitemOpen
  \bibfield  {author} {\bibinfo {author} {\bibfnamefont {S.}~\bibnamefont
  {Barraza-Lopez}}, \bibinfo {author} {\bibfnamefont {T.~P.}\ \bibnamefont
  {Kaloni}}, \bibinfo {author} {\bibfnamefont {S.~P.}\ \bibnamefont {Poudel}},
  \ and\ \bibinfo {author} {\bibfnamefont {P.}~\bibnamefont {Kumar}},\
  }\href@noop {} {\bibfield  {journal} {\bibinfo  {journal} {Phys. Rev. B}\
  }\textbf {\bibinfo {volume} {97}},\ \bibinfo {pages} {024110} (\bibinfo
  {year} {2018})}\BibitemShut {NoStop}%
\bibitem [{\citenamefont {Gomes}\ and\ \citenamefont
  {Carvalho}(2015)}]{Gomes2015}%
  \BibitemOpen
  \bibfield  {author} {\bibinfo {author} {\bibfnamefont {L.~C.}\ \bibnamefont
  {Gomes}}\ and\ \bibinfo {author} {\bibfnamefont {A.}~\bibnamefont
  {Carvalho}},\ }\href@noop {} {\bibfield  {journal} {\bibinfo  {journal}
  {Phys. Rev. B}\ }\textbf {\bibinfo {volume} {92}},\ \bibinfo {pages} {085406}
  (\bibinfo {year} {2015})}\BibitemShut {NoStop}%
\bibitem [{\citenamefont {Naumis}\ \emph {et~al.}(2017)\citenamefont {Naumis},
  \citenamefont {Barraza-Lopez}, \citenamefont {Oliva-Leyva},\ and\
  \citenamefont {Terrones}}]{Naumis2017}%
  \BibitemOpen
  \bibfield  {author} {\bibinfo {author} {\bibfnamefont {G.~G.}\ \bibnamefont
  {Naumis}}, \bibinfo {author} {\bibfnamefont {S.}~\bibnamefont
  {Barraza-Lopez}}, \bibinfo {author} {\bibfnamefont {M.}~\bibnamefont
  {Oliva-Leyva}}, \ and\ \bibinfo {author} {\bibfnamefont {H.}~\bibnamefont
  {Terrones}},\ }\href@noop {} {\bibfield  {journal} {\bibinfo  {journal}
  {Reports on Progress in Physics}\ }\textbf {\bibinfo {volume} {80}},\
  \bibinfo {pages} {096501} (\bibinfo {year} {2017})}\BibitemShut {NoStop}%
\bibitem [{\citenamefont {Hu}\ and\ \citenamefont {Kan}(2019)}]{Hu2019}%
  \BibitemOpen
  \bibfield  {author} {\bibinfo {author} {\bibfnamefont {T.}~\bibnamefont
  {Hu}}\ and\ \bibinfo {author} {\bibfnamefont {E.}~\bibnamefont {Kan}},\
  }\href@noop {} {\bibfield  {journal} {\bibinfo  {journal} {Wiley
  Interdisciplinary Reviews: Computational Molecular Science}\ }\textbf
  {\bibinfo {volume} {0}},\ \bibinfo {pages} {e1409} (\bibinfo {year}
  {2019})}\BibitemShut {NoStop}%
\bibitem [{\citenamefont {Young}\ and\ \citenamefont
  {Rappe}(2012)}]{Young2012}%
  \BibitemOpen
  \bibfield  {author} {\bibinfo {author} {\bibfnamefont {S.~M.}\ \bibnamefont
  {Young}}\ and\ \bibinfo {author} {\bibfnamefont {A.~M.}\ \bibnamefont
  {Rappe}},\ }\href@noop {} {\bibfield  {journal} {\bibinfo  {journal} {Phys.
  Rev. Lett.}\ }\textbf {\bibinfo {volume} {109}},\ \bibinfo {pages} {116601}
  (\bibinfo {year} {2012})}\BibitemShut {NoStop}%
\bibitem [{\citenamefont {Gonze}\ \emph {et~al.}(2009)\citenamefont {Gonze}
  \emph {et~al.}}]{Gonze2009}%
  \BibitemOpen
  \bibfield  {author} {\bibinfo {author} {\bibfnamefont {X.}~\bibnamefont
  {Gonze}} \emph {et~al.},\ }\href@noop {} {\bibfield  {journal} {\bibinfo
  {journal} {Computer Physics Communications}\ }\textbf {\bibinfo {volume}
  {180}},\ \bibinfo {pages} {2582} (\bibinfo {year} {2009})}\BibitemShut
  {NoStop}%
\bibitem [{\citenamefont {Perdew}\ \emph {et~al.}(1996)\citenamefont {Perdew},
  \citenamefont {Burke},\ and\ \citenamefont {Wang}}]{Perdew1996}%
  \BibitemOpen
  \bibfield  {author} {\bibinfo {author} {\bibfnamefont {J.~P.}\ \bibnamefont
  {Perdew}}, \bibinfo {author} {\bibfnamefont {K.}~\bibnamefont {Burke}}, \
  and\ \bibinfo {author} {\bibfnamefont {Y.}~\bibnamefont {Wang}},\ }\href@noop
  {} {\bibfield  {journal} {\bibinfo  {journal} {Phys. Rev. B}\ }\textbf
  {\bibinfo {volume} {54}},\ \bibinfo {pages} {16533} (\bibinfo {year}
  {1996})}\BibitemShut {NoStop}%
\bibitem [{\citenamefont {Hartwigsen}\ \emph {et~al.}(1998)\citenamefont
  {Hartwigsen}, \citenamefont {Goedecker},\ and\ \citenamefont
  {Hutter}}]{Hartwigsen1998}%
  \BibitemOpen
  \bibfield  {author} {\bibinfo {author} {\bibfnamefont {C.}~\bibnamefont
  {Hartwigsen}}, \bibinfo {author} {\bibfnamefont {S.}~\bibnamefont
  {Goedecker}}, \ and\ \bibinfo {author} {\bibfnamefont {J.}~\bibnamefont
  {Hutter}},\ }\href@noop {} {\bibfield  {journal} {\bibinfo  {journal} {Phys.
  Rev. B}\ }\textbf {\bibinfo {volume} {58}},\ \bibinfo {pages} {3641}
  (\bibinfo {year} {1998})}\BibitemShut {NoStop}%
\bibitem [{tin()}]{tiniba}%
  \BibitemOpen
  \href {https://github.com/bemese/tiniba} {}\bibinfo {note} {TINIBA is a tool
  written in bash, perl, and fortran to compute optical responses based on the
  ABINIT. https://github.com/bemese/tiniba}\BibitemShut {NoStop}%
\bibitem [{\citenamefont {Bl\"ochl}\ \emph {et~al.}(1994)\citenamefont
  {Bl\"ochl}, \citenamefont {Jepsen},\ and\ \citenamefont
  {Andersen}}]{bloch-tetra}%
  \BibitemOpen
  \bibfield  {author} {\bibinfo {author} {\bibfnamefont {P.~E.}\ \bibnamefont
  {Bl\"ochl}}, \bibinfo {author} {\bibfnamefont {O.}~\bibnamefont {Jepsen}}, \
  and\ \bibinfo {author} {\bibfnamefont {O.~K.}\ \bibnamefont {Andersen}},\
  }\href@noop {} {\bibfield  {journal} {\bibinfo  {journal} {Phys. Rev. B}\
  }\textbf {\bibinfo {volume} {49}},\ \bibinfo {pages} {16223} (\bibinfo {year}
  {1994})}\BibitemShut {NoStop}%
\bibitem [{\citenamefont {Laman}\ \emph {et~al.}(1999)\citenamefont {Laman},
  \citenamefont {Shkrebtii}, \citenamefont {Sipe},\ and\ \citenamefont {van
  Driel}}]{Laman1999}%
  \BibitemOpen
  \bibfield  {author} {\bibinfo {author} {\bibfnamefont {N.}~\bibnamefont
  {Laman}}, \bibinfo {author} {\bibfnamefont {A.~I.}\ \bibnamefont
  {Shkrebtii}}, \bibinfo {author} {\bibfnamefont {J.~E.}\ \bibnamefont {Sipe}},
  \ and\ \bibinfo {author} {\bibfnamefont {H.~M.}\ \bibnamefont {van Driel}},\
  }\href@noop {} {\bibfield  {journal} {\bibinfo  {journal} {Applied Physics
  Letters}\ }\textbf {\bibinfo {volume} {75}},\ \bibinfo {pages} {2581}
  (\bibinfo {year} {1999})}\BibitemShut {NoStop}%
\bibitem [{\citenamefont {Schmidt}\ \emph {et~al.}(1997)\citenamefont
  {Schmidt}, \citenamefont {Blanton}, \citenamefont {Hines},\ and\
  \citenamefont {Guyot-Sionnest}}]{Schmidt1997}%
  \BibitemOpen
  \bibfield  {author} {\bibinfo {author} {\bibfnamefont {M.~E.}\ \bibnamefont
  {Schmidt}}, \bibinfo {author} {\bibfnamefont {S.~A.}\ \bibnamefont
  {Blanton}}, \bibinfo {author} {\bibfnamefont {M.~A.}\ \bibnamefont {Hines}},
  \ and\ \bibinfo {author} {\bibfnamefont {P.}~\bibnamefont {Guyot-Sionnest}},\
  }\href@noop {} {\bibfield  {journal} {\bibinfo  {journal} {The Journal of
  Chemical Physics}\ }\textbf {\bibinfo {volume} {106}},\ \bibinfo {pages}
  {5254} (\bibinfo {year} {1997})}\BibitemShut {NoStop}%
\bibitem [{\citenamefont {Nastos}\ and\ \citenamefont
  {Sipe}(2010)}]{Nastos2010}%
  \BibitemOpen
  \bibfield  {author} {\bibinfo {author} {\bibfnamefont {F.}~\bibnamefont
  {Nastos}}\ and\ \bibinfo {author} {\bibfnamefont {J.~E.}\ \bibnamefont
  {Sipe}},\ }\href@noop {} {\bibfield  {journal} {\bibinfo  {journal} {Phys.
  Rev. B}\ }\textbf {\bibinfo {volume} {82}},\ \bibinfo {pages} {235204}
  (\bibinfo {year} {2010})}\BibitemShut {NoStop}%
\bibitem [{\citenamefont {Laman}\ \emph {et~al.}(2005)\citenamefont {Laman},
  \citenamefont {Bieler},\ and\ \citenamefont {van Driel}}]{Laman2005}%
  \BibitemOpen
  \bibfield  {author} {\bibinfo {author} {\bibfnamefont {N.}~\bibnamefont
  {Laman}}, \bibinfo {author} {\bibfnamefont {M.}~\bibnamefont {Bieler}}, \
  and\ \bibinfo {author} {\bibfnamefont {H.~M.}\ \bibnamefont {van Driel}},\
  }\href@noop {} {\bibfield  {journal} {\bibinfo  {journal} {Journal of Applied
  Physics}\ }\textbf {\bibinfo {volume} {98}},\ \bibinfo {pages} {103507}
  (\bibinfo {year} {2005})}\BibitemShut {NoStop}%
\bibitem [{\citenamefont {Arzate}\ \emph {et~al.}(2016)\citenamefont {Arzate},
  \citenamefont {Mendoza}, \citenamefont {V\'azquez-Nava}, \citenamefont
  {Ibarra-Borja},\ and\ \citenamefont {\'Alvarez-N\'u\~nez}}]{Arzate2016}%
  \BibitemOpen
  \bibfield  {author} {\bibinfo {author} {\bibfnamefont {N.}~\bibnamefont
  {Arzate}}, \bibinfo {author} {\bibfnamefont {B.~S.}\ \bibnamefont {Mendoza}},
  \bibinfo {author} {\bibfnamefont {R.~A.}\ \bibnamefont {V\'azquez-Nava}},
  \bibinfo {author} {\bibfnamefont {Z.}~\bibnamefont {Ibarra-Borja}}, \ and\
  \bibinfo {author} {\bibfnamefont {M.~I.}\ \bibnamefont
  {\'Alvarez-N\'u\~nez}},\ }\href@noop {} {\bibfield  {journal} {\bibinfo
  {journal} {Phys. Rev. B}\ }\textbf {\bibinfo {volume} {93}},\ \bibinfo
  {pages} {115433} (\bibinfo {year} {2016})}\BibitemShut {NoStop}%
\bibitem [{\citenamefont {Li}\ \emph {et~al.}(2018)\citenamefont {Li},
  \citenamefont {Kushnir}, \citenamefont {Wang}, \citenamefont {Dong},
  \citenamefont {Chertopalov}, \citenamefont {Rao}, \citenamefont {Mochalin},
  \citenamefont {Podila}, \citenamefont {Koski},\ and\ \citenamefont
  {Titova}}]{Li2018}%
  \BibitemOpen
  \bibfield  {author} {\bibinfo {author} {\bibfnamefont {G.}~\bibnamefont
  {Li}}, \bibinfo {author} {\bibfnamefont {K.}~\bibnamefont {Kushnir}},
  \bibinfo {author} {\bibfnamefont {M.}~\bibnamefont {Wang}}, \bibinfo {author}
  {\bibfnamefont {Y.}~\bibnamefont {Dong}}, \bibinfo {author} {\bibfnamefont
  {S.}~\bibnamefont {Chertopalov}}, \bibinfo {author} {\bibfnamefont {A.~M.}\
  \bibnamefont {Rao}}, \bibinfo {author} {\bibfnamefont {V.~N.}\ \bibnamefont
  {Mochalin}}, \bibinfo {author} {\bibfnamefont {R.}~\bibnamefont {Podila}},
  \bibinfo {author} {\bibfnamefont {K.}~\bibnamefont {Koski}}, \ and\ \bibinfo
  {author} {\bibfnamefont {L.~V.}\ \bibnamefont {Titova}},\ }in\ \href@noop {}
  {\emph {\bibinfo {booktitle} {2018 43rd International Conference on Infrared,
  Millimeter, and Terahertz Waves (IRMMW-THz)}}}\ (\bibinfo {year}
  {2018})\BibitemShut {NoStop}%
\bibitem [{\citenamefont {Pagliaro}\ \emph {et~al.}(2008)\citenamefont
  {Pagliaro}, \citenamefont {Palmisano},\ and\ \citenamefont
  {Ciriminna}}]{Pagliaro2008}%
  \BibitemOpen
  \bibfield  {author} {\bibinfo {author} {\bibfnamefont {M.}~\bibnamefont
  {Pagliaro}}, \bibinfo {author} {\bibfnamefont {G.}~\bibnamefont {Palmisano}},
  \ and\ \bibinfo {author} {\bibfnamefont {R.}~\bibnamefont {Ciriminna}},\
  }\href@noop {} {\emph {\bibinfo {title} {Flexible Solar Cells}}}\ (\bibinfo
  {publisher} {John Wiley \& Sons, Ltd},\ \bibinfo {year} {2008})\BibitemShut
  {NoStop}%
\bibitem [{\citenamefont {Tan}\ \emph {et~al.}(2016)\citenamefont {Tan},
  \citenamefont {Zheng}, \citenamefont {Young}, \citenamefont {Wang},
  \citenamefont {Liu},\ and\ \citenamefont {Rappe}}]{Tan2016}%
  \BibitemOpen
  \bibfield  {author} {\bibinfo {author} {\bibfnamefont {L.~Z.}\ \bibnamefont
  {Tan}}, \bibinfo {author} {\bibfnamefont {F.}~\bibnamefont {Zheng}}, \bibinfo
  {author} {\bibfnamefont {S.~M.}\ \bibnamefont {Young}}, \bibinfo {author}
  {\bibfnamefont {F.}~\bibnamefont {Wang}}, \bibinfo {author} {\bibfnamefont
  {S.}~\bibnamefont {Liu}}, \ and\ \bibinfo {author} {\bibfnamefont {A.~M.}\
  \bibnamefont {Rappe}},\ }\href@noop {} {\bibfield  {journal} {\bibinfo
  {journal} {npj Comput Mater}\ }\textbf {\bibinfo {volume} {2}},\ \bibinfo
  {pages} {16026} (\bibinfo {year} {2016})}\BibitemShut {NoStop}%
\bibitem [{\citenamefont {King-Smith}\ and\ \citenamefont
  {Vanderbilt}(1993)}]{King-Smith1993}%
  \BibitemOpen
  \bibfield  {author} {\bibinfo {author} {\bibfnamefont {R.~D.}\ \bibnamefont
  {King-Smith}}\ and\ \bibinfo {author} {\bibfnamefont {D.}~\bibnamefont
  {Vanderbilt}},\ }\href@noop {} {\bibfield  {journal} {\bibinfo  {journal}
  {Phys. Rev. B}\ }\textbf {\bibinfo {volume} {47}},\ \bibinfo {pages} {1651}
  (\bibinfo {year} {1993})}\BibitemShut {NoStop}%
\bibitem [{\citenamefont {Resta}(1994)}]{Resta1994}%
  \BibitemOpen
  \bibfield  {author} {\bibinfo {author} {\bibfnamefont {R.}~\bibnamefont
  {Resta}},\ }\href@noop {} {\bibfield  {journal} {\bibinfo  {journal} {Rev.
  Mod. Phys.}\ }\textbf {\bibinfo {volume} {66}},\ \bibinfo {pages} {899}
  (\bibinfo {year} {1994})}\BibitemShut {NoStop}%
\bibitem [{\citenamefont {Wang}\ and\ \citenamefont {Qian}(2017)}]{Wang2017}%
  \BibitemOpen
  \bibfield  {author} {\bibinfo {author} {\bibfnamefont {H.}~\bibnamefont
  {Wang}}\ and\ \bibinfo {author} {\bibfnamefont {X.}~\bibnamefont {Qian}},\
  }\href@noop {} {\bibfield  {journal} {\bibinfo  {journal} {2D Materials}\
  }\textbf {\bibinfo {volume} {4}},\ \bibinfo {pages} {015042} (\bibinfo {year}
  {2017})}\BibitemShut {NoStop}%
\bibitem [{\citenamefont {Wang}\ and\ \citenamefont {Qian}(2019)}]{Wang2019}%
  \BibitemOpen
  \bibfield  {author} {\bibinfo {author} {\bibfnamefont {H.}~\bibnamefont
  {Wang}}\ and\ \bibinfo {author} {\bibfnamefont {X.}~\bibnamefont {Qian}},\
  }\href@noop {} {\bibfield  {journal} {\bibinfo  {journal} {Sci. Adv.}\
  }\textbf {\bibinfo {volume} {5}},\ \bibinfo {pages} {eaav9743} (\bibinfo
  {year} {2019})}\BibitemShut {NoStop}%
\bibitem [{\citenamefont {Singh}\ and\ \citenamefont
  {Hennig}(2014)}]{Singh2014}%
  \BibitemOpen
  \bibfield  {author} {\bibinfo {author} {\bibfnamefont {A.~K.}\ \bibnamefont
  {Singh}}\ and\ \bibinfo {author} {\bibfnamefont {R.~G.}\ \bibnamefont
  {Hennig}},\ }\href@noop {} {\bibfield  {journal} {\bibinfo  {journal}
  {Applied Physics Letters}\ }\textbf {\bibinfo {volume} {105}},\ \bibinfo
  {pages} {042103} (\bibinfo {year} {2014})}\BibitemShut {NoStop}%
\bibitem [{\citenamefont {Fei}\ \emph {et~al.}(2016)\citenamefont {Fei},
  \citenamefont {Kang},\ and\ \citenamefont {Yang}}]{Fei2016}%
  \BibitemOpen
  \bibfield  {author} {\bibinfo {author} {\bibfnamefont {R.}~\bibnamefont
  {Fei}}, \bibinfo {author} {\bibfnamefont {W.}~\bibnamefont {Kang}}, \ and\
  \bibinfo {author} {\bibfnamefont {L.}~\bibnamefont {Yang}},\ }\href@noop {}
  {\bibfield  {journal} {\bibinfo  {journal} {Phys. Rev. Lett.}\ }\textbf
  {\bibinfo {volume} {117}},\ \bibinfo {pages} {097601} (\bibinfo {year}
  {2016})}\BibitemShut {NoStop}%
\bibitem [{\citenamefont {Bena}\ and\ \citenamefont
  {Montambaux}(2009)}]{Bena2009}%
  \BibitemOpen
  \bibfield  {author} {\bibinfo {author} {\bibfnamefont {C.}~\bibnamefont
  {Bena}}\ and\ \bibinfo {author} {\bibfnamefont {G.}~\bibnamefont
  {Montambaux}},\ }\href@noop {} {\bibfield  {journal} {\bibinfo  {journal}
  {New J. Phys.}\ }\textbf {\bibinfo {volume} {11}},\ \bibinfo {pages} {095003}
  (\bibinfo {year} {2009})}\BibitemShut {NoStop}%
\end{thebibliography}

%

\end{document}